\documentclass[11pt]{article}
\usepackage[utf8]{inputenc}
\usepackage[T1]{fontenc}
\pdfoutput = 1

\usepackage{subfiles}
\usepackage{graphicx}
\usepackage{listings}
\usepackage{amsthm}
\usepackage{amsmath}
\usepackage{amssymb}
\usepackage{amsfonts}
\usepackage{braket}
\usepackage{bbm}
\usepackage{bbold}
\usepackage{float}
\usepackage{mathabx}
\usepackage{url}
\usepackage{upgreek}
\usepackage{slashed}
\usepackage[normalem]{ulem}

\usepackage[hyperfootnotes = true, colorlinks = true, linkcolor = darkblue, citecolor = purple]{hyperref}

\usepackage[margin = 2.225cm]{geometry}
\usepackage{mathtools}
\allowdisplaybreaks
\numberwithin{equation}{section}

\usepackage{color}
    \definecolor{darkgreen}{rgb}{0,0.5,0}
    \definecolor{darkred}{rgb}{0.5,0,0}
    \definecolor{darkblue}{rgb}{0,0,0.6}
    \definecolor{purple}{rgb}{0.4,.2,0.7}

\usepackage[ragged]{footmisc}
\usepackage{dsfont}

\newcommand{\dd}{{\mathrm{d}}}
\newcommand{\Tr}{{\mathrm{Tr}}}
\newcommand{\hochkomma}{$^{,}$}
\newcommand{\Ai}{{\text{Ai}}}
\usepackage{mathrsfs}

\newcommand{\nb}{\text{F}}
\newcommand\TT[2]{\mathcal{T}^{#1}_{#2}}

\title{\boldmath The complete non-perturbative partition function of minimal superstring theory and JT supergravity}

\begin{document}

\thispagestyle{empty}
\begin{center}
    ~\vspace{5mm}
    
    {\Large \bf 

        The complete non-perturbative partition function of minimal superstring theory and JT supergravity
    
    }
    
    \vspace{0.4in}
    
    {\bf Dan Stefan Eniceicu,$^{1}$ Chitraang Murdia,$^{2}$ and Andrii Torchylo.$^{1}$}

    \vspace{0.4in}

    $^1$ Department of Physics, Stanford University, Stanford, CA 94305-4060, USA \vskip1ex
    $^2$ Department of Physics and Astronomy, University of Pennsylvania,\\Philadelphia, PA 19104, USA
    \vspace{0.1in}
    
    {\tt eniceicu@stanford.edu, murdia@sas.upenn.edu, torchylo@stanford.edu}
\end{center}

\vspace{0.4in}

\begin{abstract}
We derive an exact convergent expression for the partition function of the $\mathcal{N}=1$ $(2,4k)$ minimal superstring theory with type 0B GSO projection in the ungapped phase by leveraging the duality between this theory and a double-scaled unitary matrix integral. Taking the $k\rightarrow\infty$ limit, we also obtain the complete partition function of $\mathcal{N}=1$ JT supergravity, including all contributions associated with ``doubly non-perturbative'' effects. We discover that the fundamental objects of the string theory are a linear combination of the standard FZZT branes which we call F-branes, along with their charge-conjugate partners which we call anti-F-branes. Summing over the disk and cylinder diagram contributions of the F-branes and anti-F-branes and integrating over their moduli space completely reproduces our expression for the partition function from the matrix integral side of the duality. We show that the string theory can be expressed precisely in the formalism of dressed free fermions and we propose a Hilbert space interpretation of our results. We present exact expressions for the matrix integral correlators of the double-scaled eigenvalue density. 
\end{abstract}

\pagebreak

\tableofcontents

\addtocontents{toc}{\protect\setcounter{tocdepth}{2}}

\section{Introduction and summary}

Although perturbative worldsheet techniques have led to significant insight into the structure of string theory, a fully satisfying non-perturbative description of the theory has remained elusive. An attempt to make progress toward answering this question for a class of non-critical theories called minimal string theories is provided by the conjectured duality between these models and particular limits of matrix integrals \cite{Ginsparg:1993is,DiFrancesco:1993cyw}. In the simplest incarnation of this duality, one begins by considering the integral over Hermitian matrices of size $N\times N$, weighted by the exponential of a single-trace potential,
\begin{equation}
    Z(N,t_n)=\int\dd M\,\exp\left[-N\,\Tr\,V(M)\right]\,,\label{eqn:Hermitian_matrix_integral}
\end{equation}
where the potential $V(z)$ is a polynomial whose term of lowest degree is $z^2/2$. Treating the higher-order terms in $V(z)$ as interactions,
\begin{equation}
    V(z)=\frac{z^2}{2}+V_{\text{int}}(z)\,,\qquad\qquad\qquad V_{\text{int}}(z)\equiv\sum_{n=3}^d \frac{t_n}{n}z^n\,,
\end{equation}
and performing the perturbative large-$N$ expansion of the matrix integral (\ref{eqn:Hermitian_matrix_integral}), one obtains a series which can be interpreted as the exponential of the sum over tilings of closed surfaces by polygons with at most $d$ sides. Each of these tilings is weighted by an appropriate power of $1/N$, corresponding to the genus of the surface, and an appropriate power of each of the coupling constants $t_n$, corresponding to the number of polygons with $n$ sides involved in the tiling. Thus, one finds that the matrix integral (\ref{eqn:Hermitian_matrix_integral}) acts as the generating function for tilings of closed surfaces, with $1/N$ acting as the genus counting parameter.\footnote{For a comprehensive introduction to this topic, see Ref.~\cite{Eynard:2016yaa}.}\\\\
The connection with string theory appears when considering a particular limit of the terms in this expansion. This limit is called the ``double-scaling'' limit and involves taking the size $N$ of the matrix large and tuning the coupling constants $t_n$ appropriately, such that one obtains an asymptotic series in a single parameter $g_s$. In this limit, the free energy admits a perturbative genus expansion of the form,
\begin{equation}
    F(g_s)\equiv\log Z(g_s) \sim \sum_{g=0}^\infty c_g\,g_s^{2g-2}\,,
\label{eqn:F_series}
\end{equation}
where the coefficient $c_g$ matches the contribution of the closed genus $g$ surface in the dual string theory, and $g_s$ is identified with the closed string coupling constant.\\\\
Importantly, all of the statements referenced so far are perturbative. This is unfortunate since it is expected that the genus expansion (\ref{eqn:F_series}) does not converge due to the factorial growth of its coefficients. To obtain the precise function whose small-$g_s$ asymptotic expansion matches this series, one can attempt to perform Borel resummation. In this procedure, the coefficients $c_g$ are rescaled in such a way that the resulting series is convergent. One can then attempt to analytically continue the resulting series on the positive real axis and perform a Laplace transform to obtain the desired function. In most cases, including that of the minimal string theories, this procedure does not work since the function one obtains by analytically continuing the modified series has poles on the real axis. These poles are associated with instanton effects whose contributions need to be appropriately included in order to obtain the non-perturbative completion of the series.
\\\\
The set of techniques designed to do so goes by the name of resurgence\footnote{For an introduction to resurgence, see Refs.~\cite{Aniceto:2018bis, Dorigoni:2014hea, sauzin2014introduction1summabilityresurgence}.} and involves promoting the asymptotic series of the double-scaled matrix integral to a transseries \cite{Marino:2008ya,Aniceto:2011nu,Schiappa:2013opa,Aniceto:2013fka,Couso-Santamaria:2013kmu,Nishigaki:2014ija,Couso-Santamaria:2014iia,Couso-Santamaria:2015hva,Couso-Santamaria:2016vwq,Codesido:2017jwp,Ahmed:2017lhl,Gregori:2021tvs,Baldino:2022aqm,Marino:2023nem,Iwaki:2023cek,Marino:2023gxy,Eynard:2023qdr,Eynard:2023anx,Vega:2023mpu,Alexandrov:2023wdj,Marino:2024uco,Marino:2024yme},\footnote{Here, $F_0=c_0g_s^{-2}$ is the free energy contribution associated with the sphere diagram which we choose to isolate.}
%\\cite{Dorigoni:2014hea} \cite{Marino:2008ya,Aniceto:2011nu,Hatsuda:2013oxa,Nishigaki:2014ija,Ahmed:2017lhl},
\begin{equation}
    Z(g_s)\sim e^{F_0} \sum_{n=0}^\infty\sigma^n e^{-nA/g_s}Z^{(n)}(g_s) \,,
\end{equation}
and performing a generalization of the procedure outlined previously called Borel-\'Ecalle resummation. Here, $Z^{(n)}(g_s)$ is an asymptotic series associated with the $n$-instanton sector, which can be interpreted in the string theory as the sum over the contributions of the diagrams supported by a stack of $n$ instanton branes, called ZZ branes \cite{Zamolodchikov:2001ah}. On the matrix integral side of the duality, the terms multiplying the factor $\exp(-nA/g_s)$ are those obtained from the perturbative expansion of the integration over the steepest-descent contours for $n$ eigenvalue instantons. Recent progress has shown that these non-perturbative contributions match across the duality \cite{Balthazar:2019rnh, Sen:2019qqg, Balthazar:2019ypi, Sen:2020eck, Sen:2021qdk, Eniceicu:2022nay, Eniceicu:2022dru, Chakravarty:2022cgj, Eniceicu:2022xvk, Alexandrov:2023fvb, Chakrabhavi:2024szk}.\\\\
This recent evidence prompts the question: Could the duality between these string theories and double-scaled matrix integrals somehow extend beyond perturbation theory? At first, this question might seem ill-posed since we lack a non-perturbative description both of string theory and of the double-scaled matrix integral. Concretely, it is unclear how one can obtain the double-scaling limit of an expression like (\ref{eqn:Hermitian_matrix_integral}) non-perturbatively since the double-scaling limit involves taking the size of the matrix, and therefore the number of integration variables, to be infinitely large. A complementary alternative to using resurgence is appealing to the string equation \cite{Douglas:1989ve} which provides a definition of the partition function of the theory as a solution of a differential equation \cite{Johnson:2019eik,Johnson:2020heh,Johnson:2020exp,Johnson:2020mwi,Johnson:2020lns,Johnson:2021owr,Johnson:2021rsh, Johnson:2021zuo, Johnson:2021tnl, Johnson:2022wsr, Johnson:2022pou,Johnson:2023ofr,Johnson:2024bue,Johnson:2024fkm,Johnson:2024tgg}. Both resurgence and the use of the string equation are compelling options for accessing non-perturbative aspects of the theories in any background.\footnote{In the context of the string equation method, a choice of background corresponds to a choice of boundary conditions for the differential equation. In the context of resurgence, it corresponds to a choice of transseries parameters and of the contour for performing Borel-\'Ecalle resummation. For the $(2,4k)$ minimal superstring theory, the corresponding string equation is a differential equation of order $2k$. The $(2,4k)$ theory admits $k$ different types of ZZ brane instantons \cite{Seiberg:2003nm}, meaning that the transseries consists of sectors labeled by $2k$ integers, $(n_1,\dots,n_k,m_1,\dots,m_k)$, corresponding to effects associated with a stack consisting of $n_1$ ZZ branes and $m_1$ ghost ZZ branes of the first type, $n_2$ ZZ branes and $m_2$ ghost ZZ branes of the second type, and so on. Ref.~\cite{Garoufalidis:2010ya} established that the transseries required for a complete resurgence analysis of the $(2,3)$ bosonic minimal string theory requires a sum over two nonnegative integers $n$ and $m$ corresponding to effects associated with a stack of $n$ ZZ branes and $m$ ghost ZZ branes. The statement that sectors involving ghost branes need to be included in the full transseries generalizes to all of the minimal superstring theories. However, if one is only interested in the ungapped phase of the theory, one can still perform the required resummation starting with the one-parameter transseries, so we will not address this important subtlety. For more details, see Refs.~\cite{Aniceto:2011nu, Schiappa:2013opa,Gregori:2021tvs,Baldino:2022aqm,Marino:2022rpz,Schiappa:2023ned,Eynard:2023qdr}.}
However, since previous work using each of these approaches has been a tour de force, it would be interesting to establish whether a simpler and more direct method for accessing the non-perturbative physics of minimal superstring theory exists.\\\\
Another model which has received a considerable amount of attention in part due to its duality with a double-scaled matrix integral is $\mathcal{N}=1$ JT supergravity \cite{Stanford:2019vob}. Ref.~\cite{Stanford:2019vob} suggested that, similarly to how JT gravity was argued to arise as the $k\rightarrow\infty$ limit of the family of $(2,2k+1)$ bosonic minimal string theories \cite{Saad:2019lba}, JT supergravity would arise as the large-$k$ limit of the corresponding family of minimal superstring theories. This statement was proven in Ref.~\cite{Johnson:2021owr}, which provided an explicit form for the corresponding infinite-order string equation associated with JT supergravity, and defined the model non-perturbatively through its duality with the double-scaled matrix integral.\footnote{See also Ref.~\cite{Rosso:2021orf} for related work on deformations of $\mathcal{N}=1$ JT supergravity.} In the same work, the string equation associated with JT supergravity was shown to correspond to the $k\rightarrow\infty$ limit of that of the $(2,4k)$ minimal superstring theory.\footnote{Minimal superstring theory refers to a superstring theory whose worldsheet SCFT consists of the $\mathcal{N}= 1$ super-Liouville theory, the $\mathcal{N}=1$ super-Virasoro minimal CFT, and the $bc$ and the $\beta \gamma$ ghosts.} Ref.~\cite{Johnson:2021owr} also solved an appropriate truncation of the corresponding string equation numerically to obtain the quantity $r(x)$ from which the partition function can be deduced.\footnote{See also Refs.~\cite{Johnson:2019eik,Johnson:2020heh,Johnson:2020exp,Johnson:2020mwi,Johnson:2020lns} for important prior work using the string equation to explore non-perturbative aspects of JT (super)gravity.} As is the case in the context of minimal superstring theory, while resurgence and the use of the string equation are powerful techniques which allow one to probe the non-perturbative content of the theory, finding a simpler and more direct approach would be welcome. The goal of the present work is to find such an approach which works both for minimal superstring theory and for JT supergravity.\\\\
Our focus is the duality between double-scaled unitary matrix integrals and $(2,4k)$ minimal superstring theory with type 0B GSO projection \cite{Klebanov:2003wg,Seiberg:2003nm}. Prior to double-scaling, the unitary matrix integral is given by
\begin{equation}
    Z_k(N,g,t_{l}^{\pm})=\int_{U(N)}\dd U\,\exp\left[\frac{N}{g}\Tr\,V_k(U)\right]\,,\label{eqn:unitary_matrix_int_intro}
\end{equation}
where the potential $V_k(z)$ is a Laurent polynomial,
\begin{equation}
    V_k(z)=\sum_{l=1}^k\left(\frac{t_l^+}{l}z^l+\frac{t_l^-}{l}z^{-l}\right)\,.
\end{equation}
This model has a gap-closing phase transition with the simplest version being the Gross-Witten-Wadia phase transition in the $k=1$ case \cite{Gross:1980he,Wadia:1980cp}. In the ungapped phase, the perturbative piece of the free energy takes the simple form,
\begin{equation}
    F^{(0)}(N,g,t_{l}^{\pm}) = \frac{N^2}{g^2}\sum_{l=1}^k\frac{t_l^+t_l^-}{l}\,,
\end{equation}
which implies that the corresponding genus expansion of the $0$-instanton sector is trivial. Understanding the non-perturbative aspects of this theory continues to be a challenging goal.\\\\
In this work, we show that there is a path that leads to an exact convergent expression for the complete\footnote{We emphasize that our use of the word ``complete'' is meant to indicate that the result we compute represents the non-perturbative completion associated with the ungapped phase of the collection of perturbative data of the $(2,4k)$ minimal superstring theory with type 0B GSO projection. We do not intend for the word ``complete'' to be interpreted as a suggestion of generality of our results beyond the case of the ungapped phase of this theory.} non-perturbative partition function for the ungapped phase of type 0B minimal superstring theory that avoids the need to solve a differential equation or to employ resurgence. Instead, the only input required for our method is the form of the spectral curve of the desired theory, which is the same input one needs in order to set up the standard perturbative analysis for a double-scaled matrix integral. A key step in this procedure is using the Borodin-Okounkov-Geronimo-Case theorem \cite{borodin2000fredholm, Geronimo:1979iy} to rewrite the unitary matrix integral (\ref{eqn:unitary_matrix_int_intro}) as a convergent series called the Fredholm determinant expansion. This theorem was used in Ref.~\cite{Murthy:2022ien} to derive an alternative to the giant-graviton expansion for the superconformal index of $\mathcal{N}=4$ super-Yang-Mills theory \cite{Gaiotto:2021xce}, and the expansion was further studied in Refs.~\cite{Liu:2022olj, Eniceicu:2023uvd,Ezroura:2024wmp}. The expressions for the terms of the Fredholm determinant expansion which we use in this work were derived in Ref.~\cite{Eniceicu:2023uvd} using previous observations of Refs.~\cite{tracy2013diagonal,tracy2017natural} which had found a Fredholm determinant presentation for the diagonal susceptibility of the 2D Ising model. Historically, Fredholm determinants have proven essential in examining the non-perturbative physics of a large collection of models including two-dimensional quantum gravity \cite{Johnson:2021zuo, Johnson:2022wsr, Johnson:2022pou}, topological string theory \cite{Hatsuda:2013oxa,Marino:2015ixa, Grassi:2014zfa, Codesido:2015dia, Marino:2015nla, Grassi:2017qee}, ABJM theory \cite{Marino:2011eh,Hatsuda:2012hm,Hatsuda:2012dt,Hatsuda:2013gj,Hatsuda:2015gca}, and matrix integrals \cite{Eniceicu:2023cxn, Chen:2024cvf}.\\\\
This rewriting of the matrix integral is crucial because the double-scaling limit cannot be performed directly for the original integral over the $N$ eigenvalues, as this would involve tuning the number of integration variables. However, each term in the Fredholm determinant expansion has a precise double-scaling limit. Thus, by taking the double-scaling limit term by term, we obtain the complete non-perturbative partition function of the minimal superstring theory using its dual matrix integral description. Concretely, the double-scaled partition function takes the form,
\begin{equation}
\label{eqn:freddet_intro}
    \mathcal{Z}(\kappa)=1+\sum_{m=1}^\infty \mathcal{G}^{(m)}(\kappa)\,,
\end{equation}
where $\kappa \sim g_s^{-1}$ represents the genus counting parameter one would find in a perturbative treatment. The expansion terms are given by,
\begin{multline}
    \mathcal{G}^{(m)}(\kappa)=\frac{1}{m!}\prod_{i=1}^m\int\displaylimits_{-\infty+i\epsilon/2}^{+\infty+i\epsilon/2}\frac{\dd \lambda_i}{2\pi}\,\frac{1}{m!}\prod_{j=1}^m\int\displaylimits_{-\infty-i\epsilon/2}^{+\infty-i\epsilon/2}\frac{\dd \sigma_j}{2\pi}\,\frac{\prod_{1\leq i<j\leq m}(\lambda_i-\lambda_j)^2(\sigma_i-\sigma_j)^2}{\prod_{1\leq i,j\leq m}(\lambda_i-\sigma_j)^2} \\
    \times \exp\left[\kappa\sum_{i=1}^m\left(v^+(\lambda_i)+v^-(\sigma_i)\right)\right]\,,\label{eqn:mathcalGmkappa_intro}
\end{multline}
where $v^{\pm}(\xi)$ are the double-scaled effective potentials which take the form,
\begin{equation}
    v^\pm(\xi)=\pm4(-1)^k\left[\frac{4k^2}{4k^2-1}T_{2k+1}\left(\frac{i\xi}{2k}\right)-\frac{i\xi}{2k-1}T_{2k}\left(\frac{i\xi}{2k}\right)\right]\,,
\end{equation}
for the $(2,4k)$ minimal superstring theory and the form,
\begin{equation}
    v^\pm(\xi)=\pm4i\sinh(\xi)\,,
\end{equation}
for JT supergravity. We rigorously demonstrate that for $\kappa>0$, this integral expression for $\mathcal{G}^{(m)}(\kappa)$ is convergent and real-valued.\\\\
Furthermore, our method also reveals a remarkable fact about the non-perturbative description of these string theories. In usual discussions of models whose defining worldsheet CFT involves the (super-) Liouville CFT, the two types of branes these theories possess play different roles. 
The ZZ branes \cite{Zamolodchikov:2001ah} are commonly considered fundamental, in the sense that they correspond to instanton sectors whose effects need to be included in any description of the theory which claims to not disregard the non-perturbative corrections in $1/g_s$. Conversely, the FZZT branes \cite{Fateev:2000ik, Teschner:2000md} are commonly believed to play the role of probe branes that one can choose to insert into the system to measure or modify correlation functions. However, we find that this is not the case. The role of fundamental branes is played not by the ZZ branes, but instead, by a different set of branes originally introduced and studied in Ref.~\cite{Schiappa:2023ned}, which we will refer to as \emph{F-branes}. In the boundary state formalism, these are given by the difference between an FZZT brane and its corresponding ghost partner.\footnote{The ghost brane is a brane with the same FZZT modulus but opposite RR charge. In the matrix integral description, these correspond to eigenvalues living on the ``wrong'' side of the eigenvalue cut. See Refs.~\cite{Marino:2022rpz, Schiappa:2023ned, Eniceicu:2023cxn, Chakrabhavi:2024szk} for more details.} We show that the expression (\ref{eqn:freddet_intro}) for the complete non-perturbative partition function that we found from the matrix side of the duality is exactly reproduced by the sum over all disk and annulus diagrams supported on these $\nb$-branes, followed by an integration over their moduli space. In our analysis, the ZZ branes correspond to saddle points of this integral over the moduli space of the $\nb$-branes.\footnote{This point was also emphasized in Ref.~\cite{Schiappa:2023ned}.} Interestingly, our result suggests that the higher-genus diagrams supported by the $\nb$-branes must vanish, which is in contrast to the case of the diagrams supported by the ZZ branes.\footnote{One can infer that the higher-genus diagrams supported by the ZZ branes in the case of the $(2,4)$ minimal superstring do not vanish since the corresponding coefficients in the transseries derived from the matrix side of the duality do not. See equations (4.62)-(4.67) of Ref.~\cite{Marino:2008ya}.}\\\\
Expression (\ref{eqn:mathcalGmkappa_intro}) suggests that the expansion term $\mathcal{G}^{(m)}(\kappa)$ has an additional physical interpretation as a disconnected vacuum correlator involving $m$ particles and $m$ antiparticles. Denoting the contributions from connected diagrams involving $n$ particles and $n$ antiparticles by $D_{n|n}(\kappa)$, the partition function is given by
\begin{equation}
    \mathcal{Z}(\kappa)=\prod_{n=1}^\infty\exp\left(D_{n|n}(\kappa)\right)\,,
\end{equation}
while the expansion terms $\mathcal{G}^{(m)}(\kappa)$ can be expressed as,
\begin{equation}
    \mathcal{G}^{(m)}(\kappa)=\sum_{\substack{m_1,m_2,m_3,\dots\geq0\\m_1+2m_2+3m_3+\dots=m}}\prod_{n=1}^m\frac{D_{n|n}(\kappa)^{m_n}}{m_n!}\,.
\end{equation}
We find explicit expressions for $D_{n|n}(\kappa)$ in terms of integrals over the positions of the $n$ particles and $n$ antiparticles, and in terms of integrals over their momenta.\\\\
The particle-antiparticle picture motivates the question of how one should define a Hilbert space for the quantum gravity theory. 
We find that the ungapped phase of type 0B minimal superstring theory can be described as a target space theory of free one-dimensional Weyl fermions living on the real line, and that the partition function of the string theory is given by the Witten index in this fermionic description.\\\\
Moreover, we find that the partition function can be expressed directly as a Fredholm determinant,
\begin{equation}
    \mathcal{Z}(\kappa)=\det\left(\mathbb{1}-T\right)\,.
\end{equation}
The associated operator $T$ has the following momentum space representation over the Hilbert space $L^2(\mathbb{R})$,
\begin{equation}
    \braket{p'|T|p}=\theta(p)\theta(p')\int\displaylimits_{-\infty+i\epsilon}^{+\infty+i\epsilon}\frac{\dd\lambda}{2\pi}\int\displaylimits_{-\infty-i\epsilon}^{+\infty-i\epsilon}\frac{\dd\sigma}{2\pi}\,e^{ip\lambda}e^{-ip'\sigma}\frac{i}{\lambda-\sigma}\exp\left[\kappa(v^+(\lambda)+v^-(\sigma))\right]\,.
\end{equation}
We prove that the operator $T$ is trace-class which suffices to show that the aforementioned Fredholm determinant expansion is convergent.\\\\
Lastly, these techniques can be employed to compute other observables in the theory. As an illustrative example, we compute the $n$-point correlators of the eigenvalue density at the full non-perturbative level. We find that this result can be interpreted in the dual string theory by considering the sum over all diagrams involving a probe F-boundary for each factor of the eigenvalue density, followed by taking a derivative with respect to the F-boundary modulus.
\\\\
We conclude this section with an outline of the paper. In section \ref{sec:2}, we derive the complete non-perturbative partition function of the $(2,4k)$ minimal superstring theory with type 0B GSO projection in the ungapped phase by evaluating the double-scaling limit of the dual unitary matrix integral. We also reproduce this result directly within the string theory as the sum over all contributions coming from disk and cylinder diagrams supported on $\nb$-branes and anti-$\nb$-branes. In section \ref{sec:3}, we prove important properties of the expansion terms appearing in the expression we found for the partition function. We interpret these terms as disconnected correlators, and we find the expressions of the corresponding connected correlators. In section \ref{sec:4}, we present the Hilbert space formulation of our results. We prove that the expression we found for the partition function is the Fredholm determinant associated with a trace-class operator, and is therefore convergent. We interpret our results in the language of free fermions. In section \ref{sec:5}, we compute the $n$-point correlators of the eigenvalue density at the full non-perturbative level and present their interpretation in the dual string theory. We conclude with a short discussion and some proposals for future investigation in section \ref{sec:conc}. In appendix \ref{app:FredProof}, we briefly review the derivation of the Fredholm determinant expansion for a finite-$N$ unitary matrix integral. Appendices \ref{app:pure_sugra} and \ref{app:SJT} contain details pertaining to the special cases of pure supergravity ($k=1$) and JT supergravity ($k\to\infty$), respectively.
\section{The derivation of the partition function}
\label{sec:2}

In this section, we derive the complete non-perturbative partition function of the $(2,4k)$ minimal superstring theory with type 0B GSO projection in the ungapped phase. First, we show how to implement the double-scaling limit non-perturbatively for the dual unitary matrix integral with a single-trace potential. We then interpret the matrix integral result directly in the string theory. 

\subsection{The double-scaling limit of the unitary matrix integral}
\label{sec:2.1}

Consider the unitary matrix integral with a generic single-trace potential,
\begin{equation}
    Z(N,g,t_l^\pm)=\int_{U(N)}\dd U\,\exp\left[\frac{N}{g}\Tr\,V(U)\right]\,,\label{eqn:unitary_matrix_def}
\end{equation}
where $V(z)$ is a Laurent polynomial,
\begin{equation}
    V(z)=\sum_{l=1}^k\left(\frac{t_l^+}{l}z^l+\frac{t_l^-}{l}z^{-l}\right)\,,
\end{equation}
and $t_l^\pm$ for $1\leq l\leq k$ are fixed complex parameters with $(t_l^+)^*=t_l^-$, $g \in \mathbb{R}^+$ is the 't Hooft coupling, and $N$ is a positive integer denoting the degree of the unitary group. \\\\
The large-$N$ limit of (\ref{eqn:unitary_matrix_def}) takes the form,
\begin{equation}
    Z^{(0)}(N,g,t_l^\pm)=\exp\left(\frac{N^2}{g^2}\sum_{l=1}^k\frac{t_l^+t_l^-}{l}\right)\,.\label{eqn:unitary_matrix_largeN}
\end{equation}
In other words, in the strict large-$N$ limit, we have,
\begin{equation}
    \lim_{N\rightarrow\infty}\frac{Z(N,g,t_l^\pm)}{Z^{(0)}(N,g,t_l^\pm)}=1\,.
\end{equation}
In Ref.~\cite{Murthy:2022ien}, it was shown that the ratio between the unitary matrix integral (\ref{eqn:unitary_matrix_def}) and its large-$N$ limit (\ref{eqn:unitary_matrix_largeN}) can be expressed as an \textit{exact} convergent expansion called the Fredholm determinant expansion,
\begin{equation}
    \frac{Z(N,g,t_l^\pm)}{Z^{(0)}(N,g,t_l^\pm)}=1+\sum_{m=1}^\infty G_N^{(m)}\,.\label{eqn:Fredholm_det_expansion}
\end{equation}
In Ref.~\cite{Eniceicu:2023uvd}, the terms in this expansion were shown to be given by unitary supermatrix integrals,
\begin{equation}
    G_N^{(m)}=\int_{U(m|m)}\dd W\,\text{Ber}(W)^N\,\exp\left[\frac{N}{g}\sum_{l=1}^k\left(\frac{t_l^-}{l}\text{Str}\left(W^l\right)-\frac{t_l^+}{l}\text{Str}\left(W^{-l}\right)\right)\right]\,,
\end{equation}
where $\text{Ber}(W)$ denotes the Berezinian (superdeterminant) of the supermatrix $W$, and $\text{Str}(W)$ denotes the supertrace of the supermatrix $W$. More concretely, $G_N^{(m)}$ can be expressed as an integral over $2m$ complex variables $u_i$, $1\leq i\leq m$ (eigenvalues) and $v_j$, $1\leq j\leq m$ (anti-eigenvalues), as follows:
\begin{multline}
    G_N^{(m)} =(-1)^m\frac{1}{m!}\prod_{i=1}^m\oint_{r_-}\frac{\dd u_i}{2\pi i}\,\frac{1}{m!}\prod_{j=1}^m\oint_{r_+}\frac{\dd v_j}{2\pi i}\,\frac{\prod_{1\leq i<j\leq m}(u_i-u_j)^2(v_i-v_j)^2}{\prod_{1\leq i,j\leq m}(u_i-v_j)^2}\,\prod_{i=1}^m\left(\frac{u_i}{v_i}\right)^N\\
    \times\prod_{i=1}^m\exp\left[\frac{N}{g}\sum_{l=1}^k\left(\frac{t_l^-}{l}\left(u_i^l-v_i^l\right)-\frac{t_l^+}{l}\left(u_i^{-l}-v_i^{-l}\right)\right)\right]\,. \label{eqn:original_expression_GNm}
\end{multline}
Here, the eigenvalues $u_i$ are integrated along circles of radius $r_-\rightarrow 1^-$ and the anti-eigenvalues $v_j$ are integrated along circles of radius $r_+\rightarrow1^+$, all of which are centered at the origin. The expression for $G_N^{(m)}$ can be explicitly evaluated as a power series in the parameters $t_l^\pm$ by expanding the exponential and evaluating the contour integrals using the residue theorem. In appendix \ref{app:FredProof}, we briefly review the proof of this claim.\\\\
In this work, we are interested in particular limits of (\ref{eqn:unitary_matrix_def}) which correspond to the partition functions of the $(2,4k)$ minimal superstring theories. Our main goal is to obtain an exact expression for the complete non-perturbative partition functions of these theories by using the Fredholm determinant expansion (\ref{eqn:Fredholm_det_expansion}).\\\\
The partition function of the $(2,4k)$ minimal superstring theory is obtained from (\ref{eqn:unitary_matrix_def}) by taking the double-scaling limit wherein we zoom into the region around the $k^\text{th}$ multi-critical point while taking $N\to\infty$. The $k^\text{th}$ multi-critical point corresponds to the parameters \cite{mandal1990phase,periwal1990unitary,Oota:2021qky}
\begin{equation}
    t_l^\pm = \left\{
	\begin{array}{ll}
		\frac{g (k!)^2}{(k+l)!(k-l)!}\,, & 1\leq l\leq k\,, \\
		0\,, & l>k\,.
	\end{array}
\right.
\end{equation}
The procedure of obtaining the spectral curve associated with the $(2,4k)$ minimal superstring theory was explained in Ref.~\cite{Chakrabhavi:2024szk}; we recall the main points here for convenience.\\\\
In the $N\rightarrow\infty$ limit, the leading eigenvalue distribution of the unitary matrix integral (\ref{eqn:unitary_matrix_def}) can cover either the entire unit circle (the ungapped phase) or a subset of the unit circle (the gapped phase) depending on the values of the parameter $g$ and of the couplings $t_l^\pm$. The double-scaling procedure zooms into the region of the gap-closing phase transition. We restrict to approaching the transition point exclusively from the ungapped phase.\\\\
A probe eigenvalue feels an effective potential due to the combination of the external potential $-V(z)$ and the Vandermonde repulsion from the other eigenvalues. In the ungapped phase, the effective potential takes different forms inside and outside the unit circle:
\begin{equation}
    V_{\text{eff}}^\pm(z)=\mp\left[g\log z+\sum_{l=1}^k\left(\frac{t_l^+}{l}z^l-\frac{t_l^-}{l}z^{-l}\right)\right]\,.\label{eqn:Veff}
\end{equation}
Here $V_{\text{eff}}^+$ is naturally defined outside the unit circle while $V_{\text{eff}}^-$ is naturally defined inside. The spectral curve is defined by the relation,
\begin{equation}
    y^\pm(z)=-V_{\text{eff}}^{\pm}{}'(z)=\pm\left[\frac{g}{z}+\sum_{l=1}^k\left(t_l^+z^{l-1}+t_l^-z^{-l-1}\right)\right]\,.\label{eqn:spectral_curve_matrix}
\end{equation}
The double-scaling limit is defined such that this spectral curve matches the one associated with the $(2,4k)$ minimal superstring\footnote{Here, $T_n(x)$ denotes the $n$th Chebyshev polynomial of the first kind. It is defined by the relation, $T_n(\cos(\theta))=\cos(n\theta)$.},
\begin{equation}
    \Tilde{y}^\pm(x)=\pm i(-1)^k\;T_{2k}\left(\frac{ix}{2k}\right)\,.
\end{equation}
The gap-closing transition happens at the point $z=-1$, where the two ends of the large-$N$ eigenvalue distribution in the gapped phase connect. The double-scaling limit requires zooming in near this point,\footnote{Note that the way we are taking the double-scaling limit here differs slightly from how it is taken in Ref.~\cite{Chakrabhavi:2024szk}. The coordinate $x$ is rescaled by a factor of $k$, so that we can directly take the JT supergravity ($k\rightarrow\infty$) limit without performing any additional rescaling.}
\begin{equation}
    z=-1+\frac{i\varepsilon x}{k}\,,\label{eqn:zoom}
\end{equation}
and tuning the couplings around their values at the $k$th multi-critical point in a specific way,
\begin{equation}
    t_l^\pm=\frac{g(k!)^2}{(k+l)!(k-l)!}\left(1+\sum_{m=1}^ka_{l,m}\varepsilon^{2m}\right)\,.
\end{equation}
Here, $\varepsilon$ is a positive real parameter that allows us to perform the double-scaling procedure, and $a_{l,m}$ are finite constants\footnote{It is possible to find the values of the constants $a_{l,m}$ explicitly given $k$. For instance, for $k=1$, the only nonzero constant is $a_{1,1}=-1$. For $k=2$, one can take $a_{1,1}=-\frac{1}{3}$, $a_{2,1}=-\frac{4}{3}$, and $a_{2,2}=1$ as the only nonzero constants. We will not need the precise values of the constants in this work.} which are determined by requiring that
\begin{equation}
    y^{\pm}(z)=2ig\binom{2k}{k}^{-1}\varepsilon^{2k}\Tilde{y}^\pm(x)+O(\varepsilon^{2k+1})\,.\label{eqn:spectral_curve_ds}
\end{equation}
The double-scaling limit involves taking $N\rightarrow\infty$ and $\varepsilon\rightarrow0$, while holding fixed the quantity
\begin{equation}
    \kappa=\frac{1}{2k}\binom{2k}{k}^{-1}N\varepsilon^{2k+1}\,.
\end{equation}
The effective potential in the double-scaling limit is
\begin{align}
    V_{\text{eff}}^\pm(z)&=V_{\text{eff}}^\pm(-1)-\int_{-1}^z\dd\zeta\;y^\pm(\zeta)\nonumber\\
    &=V_{\text{eff}}^\pm(-1)\pm\frac{g}{N}\times 4(-1)^k\kappa\left[\frac{4k^2}{4k^2-1}T_{2k+1}\left(\frac{ix}{2k}\right)-\frac{ix}{2k-1}T_{2k}\left(\frac{ix}{2k}\right)\right]+O(\varepsilon^{2k+2})\,.\label{eqn:expression_for_d-s_effective_potential}
\end{align}
The main observation we want to highlight in this section is that each term $G_N^{(m)}$ of the Fredholm determinant expansion (\ref{eqn:Fredholm_det_expansion}) has a well-defined double-scaling limit. This observation allows us to express the non-perturbatively complete partition function of the $(2,4k)$ minimal superstring theory as a relatively simple convergent series up to an overall constant,
\begin{equation}
    \mathcal{Z}(\kappa)=1+\sum_{m=1}^\infty \mathcal{G}^{(m)}(\kappa)\,.\label{eqn:BOGC_expansion_string}
\end{equation}
To determine the double-scaled versions of these terms, we note that near the $k$th multi-critical point,
\begin{multline}
    \left(\frac{u_i}{v_i}\right)^N\exp\left[\frac{N}{g}\sum_{l=1}^k\left(\frac{t_l^-}{l}\left(u_i^l-v_i^l\right)-\frac{t_l^+}{l}\left(u_i^{-l}-v_i^{-l}\right)\right)\right] \\
    =\exp\left[\frac{N}{g}\left(V_{\text{eff}}^+\left(\frac{1}{u_i}\right)+V_{\text{eff}}^-\left(\frac{1}{v_i}\right)\right)\right]\,.
\end{multline}
Following the relation (\ref{eqn:zoom}), we zoom in around $-1$ by taking
\begin{align}
    u_i&=\left(-1+\frac{i\varepsilon \lambda_i}{k}\right)^{-1}=-1-\frac{i\varepsilon \lambda_i}{k}+O(\varepsilon^2)\,,\\ v_i&=\left(-1+\frac{i\varepsilon \sigma_i}{k}\right)^{-1}=-1-\frac{i\varepsilon\sigma_i}{k}+O(\varepsilon^2)\,.
\end{align}
The Vandermonde superdeterminant becomes
\begin{equation}
    \frac{\prod_{1\leq i<j\leq m}(u_i-u_j)^2(v_i-v_j)^2}{\prod_{1\leq i,j\leq m}(u_i-v_j)^2}=(-1)^m\left(\frac{\epsilon}{k}\right)^{-2m}\frac{\prod_{1\leq i<j\leq m}(\lambda_i-\lambda_j)^2(\sigma_i-\sigma_j)^2}{\prod_{1\leq i,j\leq m}(\lambda_i-\sigma_j)^2}\,,
\end{equation}
and the terms in the Fredholm determinant expansion become
\begin{multline}
    \mathcal{G}^{(m)}(\kappa) =\frac{1}{m!}\prod_{i=1}^m\int_{\mathcal{C}_+}\frac{\dd \lambda_i}{2\pi}\,\frac{1}{m!}\prod_{j=1}^m\int_{\mathcal{C}_-}\frac{\dd \sigma_j}{2\pi}\,\frac{\prod_{1\leq i<j\leq m}(\lambda_i-\lambda_j)^2(\sigma_i-\sigma_j)^2}{\prod_{1\leq i,j\leq m}(\lambda_i-\sigma_j)^2} \\
    \times\exp\left[\frac{N}{g}\left(V_{\text{eff}}^+\left(\frac{1}{u_i}\right)+V_{\text{eff}}^-\left(\frac{1}{v_i}\right)\right)\right]\,.
\end{multline}
We can now use our previous result (\ref{eqn:expression_for_d-s_effective_potential}) for the double-scaled effective potential to conclude that
\begin{multline}
    \mathcal{G}^{(m)}(\kappa)=\frac{1}{m!}\prod_{i=1}^m\int_{\mathcal{C}_+}\frac{\dd \lambda_i}{2\pi}\,\frac{1}{m!}\prod_{j=1}^m\int_{\mathcal{C}_-}\frac{\dd \sigma_j}{2\pi}\,\frac{\prod_{1\leq i<j\leq m}(\lambda_i-\lambda_j)^2(\sigma_i-\sigma_j)^2}{\prod_{1\leq i,j\leq m}(\lambda_i-\sigma_j)^2} \\
    \times \exp\left[\kappa\sum_{i=1}^m\left(v^+(\lambda_i)+v^-(\sigma_i)\right)\right]\,,\label{eqn:mathcalGmkappa}
\end{multline}
where the double-scaled effective potentials are given by
\begin{equation}
\label{eqn:veff_pm}
    v^\pm(\xi)=\pm4(-1)^k\left[\frac{4k^2}{4k^2-1}T_{2k+1}\left(\frac{i\xi}{2k}\right)-\frac{i\xi}{2k-1}T_{2k}\left(\frac{i\xi}{2k}\right)\right]\,.
\end{equation}
The contours $\mathcal{C}_+$ and $\mathcal{C}_-$ should be consistent with their counterparts before the double-scaling procedure was performed. Since double scaling zooms into the region near the point $-1$, the portions of the integration contours for the $u_i$ and $v_j$ variables in this region will become straight lines parallel to the real axis, with $\mathcal{C}_+$ above the real axis and $\mathcal{C}_-$ below the real axis.\\\\
We will discuss the relevant properties of the integral expressions (\ref{eqn:mathcalGmkappa}) for the individual terms of the Fredholm determinant expansion of the partition function in the next section. For now, we summarize the main points. These expressions are convergent for any choices of parallel contours $\mathcal{C}_+$ above and $\mathcal{C}_-$ below the real axis. In fact, due to the holomorphicity of the integrand in each of the variables $\lambda_i$ and $\sigma_j$, the result of (\ref{eqn:mathcalGmkappa}) is independent of the choice of contours. Essentially, since the integrands are holomorphic in each variable and decay sufficiently rapidly at infinity, the integrals do not depend on the distances between the contours $\mathcal{C}_\pm$ and the real axis. The contours can be continuously deformed as long as they do not touch the real axis and their behavior as they approach infinity remains unchanged. A particularly convenient choice is to take
\begin{equation}
    \mathcal{C}_\pm=\mathbb{R}\pm i\epsilon/2\,,
\end{equation}
where $\epsilon$ is a positive real constant. With this choice, one can immediately conclude that the partition function (\ref{eqn:BOGC_expansion_string}) must be manifestly real since $v^\pm(\xi)^*=v^\mp(\xi^*)$.

\subsection{The minimal superstring theory partition function}

We have shown that the duality between the $\mathcal{N}=1$ $(2,4k)$ minimal superstring theory with type 0B GSO projection and a particular double-scaled unitary matrix integral allows us to make a prediction for the complete partition function of the string theory using matrix integral techniques. Here, we explain how one can obtain the same result strictly within the string theory without referencing its duality to a matrix integral. The main observation we want to highlight here is the existence of a set of fundamental branes whose contributions to the partition function exactly reproduce the expression (\ref{eqn:BOGC_expansion_string}) we found in the previous subsection.\\\\
It is known that the minimal superstring admits two types of branes: the FZZT branes, which are parametrized by a modulus $z$ associated with a point on the spectral curve, and the ZZ branes, which are similarly parametrized by a modulus $z^*$. In contrast to the continuous modulus $z$ of an FZZT brane, the modulus $z^*$ of a ZZ brane can only take a finite set of values associated with the singularities of the spectral curve.\\\\
In the ungapped phase, branes are also distinguished by their RR charges. Aside from their moduli, FZZT and ZZ branes are parametrized by a sign $\xi=\pm1$ indicating their charge. Since the spectral curves associated with the minimal superstring theories in the ungapped phase have two sheets, we can think of $z$ as a point in the complex plane,\footnote{More precisely, a point on the Riemann sphere.} with the upper half-plane naturally representing positive-charge branes, and the lower half-plane naturally representing negative-charge branes, similarly to how the effective potential $v^\pm(z)$ for the matrix integral was originally defined in the upper ($+$) or lower ($-$) half-plane.\\\\
Following Ref.~\cite{Chakrabhavi:2024szk}, we will refer to a brane as positively charged if $\xi=+1$, and as negatively charged if $\xi=-1$. We will call a brane real if the sign of the imaginary part of its modulus $z$ is the same as $\xi$. Conversely, we will refer to a brane as a ghost brane if the sign of the imaginary part of its modulus $z$ is opposite to $\xi$. We will use the term ``charge-conjugate partner'' of a brane $A$ with modulus $z$ and sign $\xi$ to denote the brane $\overline{A}$ with modulus $z^*$ and sign $-\xi$. We will use the term ``ghost partner'' of a brane $A$ with modulus $z$ and sign $\xi$ to denote the brane $A^{\text{gh}}$ with the same modulus $z$, but opposite sign $-\xi$.\\\\
Focusing on the boundary state formalism, we shall denote the ket associated with the FZZT brane with parameter $z$ and charge $\xi$ by $\ket{\text{FZZT}: (z, \xi)}$. We highlight a different type of brane which was first introduced and played a central role in Ref.~\cite{Schiappa:2023ned} in the context of bosonic minimal string theory. We will refer to this brane as an F-brane,\footnote{The label F here stands for ``fundamental'' due to their essential role in reproducing the expression for the partition function of the minimal superstring theory. Despite the naming convention, there is no connection with F-theory.} and we recall its definition \cite{Schiappa:2023ned} as the difference in the boundary state formalism between an FZZT brane and its ghost partner:
\begin{equation}
    \ket{\nb : (z, \xi)} = \ket{\text{FZZT}: (z, \xi)} - \ket{\text{FZZT}: (z, - \xi)} \, .
\end{equation}
More precisely, we will refer to a brane as an F-brane if its charge $\xi$ is positive and as an anti-F-brane if its charge $\xi$ is negative. Since the tensions of real branes and their ghost partners have opposite signs,
\begin{equation}
    \TT{\text{FZZT}}{(z, -\xi)} = - \TT{\text{FZZT}}{(z,\,\xi)} \, ,
\end{equation}
the tension of the $\nb$-brane is
\begin{equation}
    \TT{\nb}{(z,\,\xi)} =  2 \TT{\text{FZZT}}{(z,\,\xi)} \, .
\end{equation}
The tensions are only defined up to an overall $g_s$-dependent factor, so we can only match the ratio of tensions to their analogs on the dual matrix integral side. To perform this matching, we normalize using the tension of the $(m,n)=(1,1)$ ZZ brane. Note that the tension of the ZZ brane is given by
\begin{equation}
    \TT{\text{ZZ}}{((1,n),\,\pm 1)} =  2 \TT{\text{FZZT}}{ ( z^*_{(1,n)},\,1 )}  \, .
\end{equation}
Here, $z^*_{(1,n)} \in i \mathbb{R}^+$ is the value of the FZZT parameter associated with the $(1,n)$ ZZ brane with $n \in \{ 1, 3, \dots 2k - 1 \} \, $ \cite{Seiberg:2003nm, Chakrabhavi:2024szk}.
Thus, for $\nb$-branes with labels $(\lambda,+1)$ and anti-$\nb$-branes with labels $(\sigma,-1)$, respectively, we have
\begin{equation}
\begin{split}
    \frac{\TT{\nb}{(\lambda,\,1)}}{\TT{\text{ZZ}}{((1,\,1), \pm 1)}} 
    &=  \frac{ 2\TT{\text{FZZT}}{(\lambda,\,1)}}{2\TT{\text{FZZT}}{ \left( z^*_{(1,1)},\,1 \right)}} 
    = \frac{v^+(\lambda)}{v^+\left(z^*_{(1,1)} \right)} \, , \\
    \frac{\TT{\nb}{(\sigma,-1)}}{\TT{\text{ZZ}}{((1,\,1), \pm 1)}} 
    &= \frac{2\TT{\text{FZZT}}{(\sigma,-1)}}{2\TT{\text{FZZT}}{ \left( z^*_{(1,1)},\,1 \right)}} 
    = \frac{v^-(\sigma)}{v^+\left(z^*_{(1,1)} \right)} \, .
\end{split} 
\end{equation}
The tension of an FZZT brane with modulus $z$ and charge $\xi$ is identified with the string theory disk diagram supported on the FZZT brane and equals $\kappa v^{\xi}(z)/2$ with our choice of conventions \cite{Seiberg:2003nm, Chakrabhavi:2024szk}. The disk diagram supported on the corresponding (anti-)F-brane therefore equals $\kappa v^{\xi}(z)$. An F-brane with modulus $\lambda$ can support an arbitrary number of disk diagrams whose total contribution to the partition function is $\exp(\kappa v^+(\lambda))$. Similarly, an anti-F-brane with modulus $\sigma$ can support an arbitrary number of disk diagrams, whose total contribution is $\exp(\kappa v^-(\sigma))$. Thus, the contribution from all possible disk diagrams supported on $m$ F-branes with moduli $\lambda_i$ and on $m$ anti-F-branes with moduli $\sigma_i$ reproduces the exponential factor in equation (\ref{eqn:mathcalGmkappa}).\\\\
The annulus diagram between two FZZT branes is given by \cite{Okuyama:2005rn}
\footnote{The result in Ref.~\cite{Okuyama:2005rn} can be generalized to the $(2, 4k)$ case by modifying their equation (2.9) to $\theta = \frac{\pi b P}{2} = \frac{\pi P}{2 \sqrt{2 k}}$.}
\begin{equation}
    A^{\text{FZZT}}_{(z,\xi),(z',\xi')} = - \frac{1 - \xi \xi'}{4} \log \left( \xi z + \xi' z' \right)^2 \, .
\end{equation}
Hence, the annulus between two $\nb$-branes with the same charge is
\begin{equation}
    A^{\nb}_{(z,\xi), (z',\xi)} = - A^{\text{FZZT}}_{(z,\xi),(z',-\xi)} - A^{\text{FZZT}}_{(z,-\xi),(z',\xi')} = \log (z - z')^2 \, ,
\end{equation}
and the annulus between two $\nb$-branes with opposite charge is
\begin{equation}
    A^{\nb}_{(z,\xi), (z',-\xi)} = A^{\text{FZZT}}_{(z,\xi),(z',-\xi)} + A^{\text{FZZT}}_{(z,-\xi),(z',\xi')} = - \log (z - z')^2 \, .
\end{equation}
If we have $m$ $\nb$-branes with moduli $\lambda_i$ and $m$ anti-$\nb$-branes with moduli $\sigma_i$, the exponentiated sum of all annulus diagrams is
\begin{equation}
    e^{A} = \frac{\prod_{1\leq i<j\leq m}(\lambda_i-\lambda_j)^2(\sigma_i-\sigma_j)^2}{\prod_{1\leq i,j\leq m}(\lambda_i-\sigma_j)^2} \, ,
\end{equation}
which matches the Vandermonde superdeterminant in \eqref{eqn:mathcalGmkappa}.\\\\
If one assumes that the higher genus diagrams supported by the (anti-)$\nb$-branes vanish for the minimal superstring theory with type 0B GSO projection in the ungapped phase, upon integrating over the moduli space of the $m$ $\nb$-branes and of the $m$ anti-$\nb$-branes, one fully recovers the expression \eqref{eqn:mathcalGmkappa}.\footnote{One could ask why the moduli are integrated over lines above or below the real axis. One should really think of the moduli $\lambda$ and $\sigma$ as real-valued, with small imaginary parts introduced for regularization purposes.}\hochkomma\footnote{It is not clear why the diagram contributions associated with $\nb$-branes appear to be one-loop exact. This is not true for ZZ branes, and we do not have an independent string theory argument for this. However, the exact match with \eqref{eqn:mathcalGmkappa} strongly suggests that this statement is true.} The interpretation of \eqref{eqn:BOGC_expansion_string} from the point of view of the string theory is that the string theory vacuum consists of sectors with an arbitrary number $m$ of excitations in the form of (anti-)$\nb$-branes; since the vacuum is neutral, the number of $\nb$-branes must match the number of anti-$\nb$-branes in each sector. These $\nb$-branes and anti-$\nb$-branes have moduli $\lambda_i$ and $\sigma_i$, respectively, which can take any real value.\footnote{Note that one needs to introduce appropriate regularization in the moduli space integral in order to avoid poles of the form $1/(\lambda_i-\sigma_j)^2$ coming from the annulus diagrams between an $\nb$-brane with modulus $\lambda_i$ and an anti-$\nb$-brane with modulus $\sigma_j$. This issue was also highlighted and treated similarly in Ref.~\cite{Schiappa:2023ned}.}
\section{Properties of the partition function}
\label{sec:3}

In this section, we study the main properties of our expression for the complete non-perturbative partition function of the $(2,4k)$ minimal superstring theory with type 0B GSO projection in the ungapped phase,
\begin{equation}
    \mathcal{Z}(\kappa)=1+\sum_{m=1}^\infty \mathcal{G}^{(m)}(\kappa)\,.\label{eqn:partition_function_3_1}
\end{equation}
We will show later that the full partition function can be written as the Fredholm determinant associated with a particular integral operator $T$ on $L^2(\mathbb{R})$. For now, we focus instead on the explicit expressions for the terms in the series,\footnote{The convergence of the full partition function (\ref{eqn:partition_function_3_1}) will be shown in section \ref{sec:4}.}
\begin{multline}
    \mathcal{G}^{(m)}(\kappa) =\frac{1}{m!}\prod_{i=1}^m\int_{\mathcal{C}_+}\frac{\dd \lambda_i}{2\pi}\,\frac{1}{m!}\prod_{j=1}^m\int_{\mathcal{C}_-}\frac{\dd \sigma_j}{2\pi}\,\frac{\prod_{1\leq i<j\leq m}(\lambda_i-\lambda_j)^2(\sigma_i-\sigma_j)^2}{\prod_{1\leq i,j\leq m}(\lambda_i-\sigma_j)^2} \\
    \times \exp\left[\kappa\sum_{i=1}^m\left(v^+(\lambda_i)+v^-(\sigma_i)\right)\right]\,,
\end{multline}
where $v^{\pm}(\xi)$ are the double-scaled effective potentials given in (\ref{eqn:veff_pm}).
The case of JT supergravity which can be obtained by taking $k\rightarrow\infty$ in the previous expressions requires special care, and therefore, we treat it separately in appendix \ref{app:SJT}.\\\\
We will first derive an alternate expression for the terms $\mathcal{G}^{(m)}(\kappa)$ using the Cauchy determinant identity. We will then show that this expression is convergent, independent of the choice of contours $\mathcal{C}_\pm$, and real. Using an elementary combinatorial argument, we provide a physical interpretation for these terms as disconnected vacuum correlators, and express them in terms of momentum space integrals using the Fourier transform. We also obtain simple expressions for the corresponding connected vacuum correlators.

\subsection{Properties of the expansion terms}\label{sec:3.1}

In this subsection, we derive some important properties of the expansion terms such as convergence, independence of the choice of contours, and real-valuedness.
We begin by using the Cauchy determinant identity,
\begin{equation}
    \det\left[\frac{1}{\lambda_i-\sigma_j}\right]_{i,j=1,\dots,m}=(-1)^{m(m-1)/2}\frac{\prod_{1\leq i<j\leq m}\left(\lambda_i-\lambda_j\right)\left(\sigma_i-\sigma_j\right)}{\prod_{i,j=1}^m\left(\lambda_i-\sigma_j\right)}\,,
\end{equation}
to rewrite the square of the Vandermonde superdeterminant as a sum over permutations $p$ and $q$,
\begin{align}
    \frac{\prod_{1\leq i<j\leq m}(\lambda_i-\lambda_j)^2(\sigma_i-\sigma_j)^2}{\prod_{1\leq i,j\leq m}(\lambda_i-\sigma_j)^2}=\sum_{p,q\,\in\,S_m}\text{sgn}(p)\,\text{sgn}(q)\prod_{i=1}^m\frac{1}{\left(\lambda_i-\sigma_{p(i)}\right)\left(\lambda_i-\sigma_{q(i)}\right)}\,.
\end{align}
Therefore, the $m$th term in the series becomes,\footnote{Here, we are taking $\lambda_j=x_j+i\epsilon_j^+$, and $\sigma_j=y_j-i\epsilon_j^-$.}
\begin{multline}
    \mathcal{G}^{(m)}(\kappa)=\sum_{p,q\,\in\,S_m}\text{sgn}(p)\,\text{sgn}(q)\frac{1}{m!}\prod_{j=1}^m\int_{-\infty}^{\infty}\frac{\dd x_j}{2\pi}\,\frac{1}{m!}\prod_{j=1}^m\int_{-\infty}^{\infty}\frac{\dd y_j}{2\pi}\,\prod_{j=1}^m\frac{1}{x_j-y_{p(j)}+\frac{i}{2}(\epsilon_j^++\epsilon_{p(j)}^-)}\\
    \times\prod_{j=1}^m\frac{1}{x_j-y_{q(j)}+\frac{i}{2}(\epsilon_j^++\epsilon_{q(j)}^-)}\times\prod_{j=1}^m\exp\left[\kappa\; v^+\left(x_j+\frac{i\epsilon_j^+}{2}\right)+\kappa\; v^-\left(y_j-\frac{i\epsilon_j^-}{2}\right)\right]\,.
\label{eqn:ToBeAnalyzed}
\end{multline}
First, we show that this expression is finite. Let us focus on the integral associated with the variable $x_j$ for a fixed choice of permutations $p$ and $q$, and show that the integral is absolutely convergent:
\begin{equation}
    \int_{-\infty}^\infty\frac{\dd x_j}{2\pi}\left|\frac{1}{x_j-y_{p(j)}+\frac{i}{2}(\epsilon_j^++\epsilon_{p(j)}^-)}\frac{1}{x_j-y_{q(j)}+\frac{i}{2}(\epsilon_j^++\epsilon_{q(j)}^-)}\exp\left[\kappa v^+\left(x_j+\frac{i\epsilon_j^+}{2}\right)\right]\right|<\infty\,.
\end{equation}
Indeed, the denominators are bounded by an expression involving only the choice of regularization parameters $\epsilon^\pm_j$,
\begin{equation}
    \left|\frac{1}{x_j-y_{p(j)}+\frac{i}{2}(\epsilon_j^++\epsilon_{p(j)}^-)}\frac{1}{x_j-y_{q(j)}+\frac{i}{2}(\epsilon_j^++\epsilon_{q(j)}^-)}\right|\leq\frac{4}{(\epsilon_j^++\epsilon_{p(j)}^-)(\epsilon_j^++\epsilon_{q(j)}^-)}\,,\label{eqn:denominator_bound}
\end{equation}
and the exponent is a polynomial in $x_j$ whose real part has a leading-order term of the form $-\kappa\,\epsilon_j^+(x_j/k)^{2k}$. Thus, the integral associated with the variable $x_j$ is bounded in absolute value by a quantity that is independent of the other $2m-1$ variables. An analogous argument serves to show that the same statement can be made for the integral associated with the variable $y_j$. Therefore, expression (\ref{eqn:ToBeAnalyzed}) is convergent for any choice of positive constants $\epsilon_j^\pm$.\\\\
In fact, the same argument allows us to show that (\ref{eqn:ToBeAnalyzed}) is independent of the choices of regularization parameters $\epsilon^\pm_j$. One can consider the integral associated with the variable $x_j$ over the rectangle with sides of length $L$ parallel to the real axis and sides of length $\Delta\epsilon_j^+$ parallel to the imaginary axis and whose base is located at $\epsilon_j^+$. The contour integral vanishes for any choices of $L$ and $\Delta\epsilon_j^+$ since the integrand appearing in (\ref{eqn:ToBeAnalyzed}) is holomorphic in $x_j$. Thus, it is sufficient to show that the integrals over the sides parallel to the imaginary axis vanish as we take $L\rightarrow\infty$. One can use the bound in (\ref{eqn:denominator_bound}) for the denominator and bound the exponent at $x_j=\pm L+iu$ as
\begin{equation}
    \left|\exp\left[\kappa v^+\left(x_j+\frac{i\epsilon_j^+}{2}\right)\right]\right|\leq\max_{u\in[0,\Delta\epsilon_j^+]}\exp\left[\kappa\;\text{Re}\left[v^+\left(L+\frac{i}{2}(\epsilon_j^++u)\right)\right]\right]\,.
\end{equation}
The leading-order term in the real part of the exponent takes the form $-\kappa (\epsilon_j^++u)(L/k)^{2k}$, which vanishes as $L\rightarrow\infty$, meaning that the integrals over the sides of the rectangle that are parallel to the imaginary axis vanish as $L\rightarrow\infty$. Thus, the integral over $x_j$ appearing in (\ref{eqn:ToBeAnalyzed}) is independent of the choice of regularization parameter, $\epsilon_j^+>0$. An analogous argument shows that (\ref{eqn:ToBeAnalyzed}) is also independent of the regularization parameters, $\epsilon_j^->0$, associated with the variables $y_j$.\footnote{The same statement holds for JT supergravity, but only for regularization parameters satisfying $0<\epsilon_j^\pm<2\pi$ due to the periodicity of the hyperbolic cosine in the imaginary part of its argument. See appendix \ref{app:SJT} for details.}\\\\
Finally, setting all regularization parameters to be equal, $\epsilon_j^+=\epsilon_j^-=\epsilon>0$, we have
\begin{multline}
    \mathcal{G}^{(m)}(\kappa)=\sum_{p,q\,\in\,S_m}\text{sgn}(p)\,\text{sgn}(q)\frac{1}{m!}\prod_{j=1}^m\int_{-\infty}^{\infty}\frac{\dd x_j}{2\pi}\,\frac{1}{m!}\prod_{j=1}^m\int_{-\infty}^{\infty}\frac{\dd y_j}{2\pi}\,\prod_{j=1}^m\frac{1}{x_j-y_{p(j)}+i\epsilon} \\
    \times\prod_{j=1}^m\frac{1}{x_j-y_{q(j)}+i\epsilon}\times\prod_{j=1}^m\exp\left[\kappa\; v^+\left(x_j+\frac{i\epsilon}{2}\right)+\kappa\; v^-\left(y_j-\frac{i\epsilon}{2}\right)\right]\,.
\label{eqn:Gm_same_epsilon}
\end{multline}
Taking the complex conjugate and relabeling the appropriate variables, one sees that this expression is manifestly real, as was argued in section \ref{sec:2.1}.

\subsection{The expansion terms as disconnected diagrams}

In this subsection, we show that the expansion term $\mathcal{G}^{(m)}(\kappa)$ can be interpreted as the sum over (possibly disconnected) bubble diagrams involving $m$ particles and $m$ antiparticles. We will later identify these particles as dressed free fermions in section \ref{sec:4}.\\\\
The partition function receives contributions from all possible particle-antiparticle creation-annihilation events occurring in the vacuum. Charge conservation requires that an event contain an equal number of particles and antiparticles, but each event can consist of any number of independent connected diagrams, with each connected diagram representing the production of a subset of particles and an equal number of antiparticles. Each connected diagram respects momentum conservation among its constituents through the presence of a Dirac delta function in its momentum-space representation. The partition function of the superstring theory consists of a sum over contributions from each such event, with various ways of organizing the diagram combinatorics.\\\\
There are two choices that are particularly useful. The first is the usual presentation of the partition function as a product of the exponentials of the contributions from connected diagrams involving $n$ particles and $n$ antiparticles, which we denote by  $D_{n|n}(\kappa)$. It follows that the partition function can be written as,
\begin{equation}
    \mathcal{Z}(\kappa)=\prod_{n=1}^\infty\exp\left(D_{n|n}(\kappa)\right)\,.\label{eqn:partition_function_as_sum_over_connected}
\end{equation}
The free energy, which is defined as the logarithm of the partition function, then represents the sum over all possible connected diagrams consisting of an equal number of particles and antiparticles,
\begin{equation}
    F(\kappa)\equiv\log\mathcal{Z}(\kappa)=\sum_{n=1}^\infty D_{n|n}(\kappa)\,.
\end{equation}
The second presentation of the partition function which is particularly useful is the one where we sum over all possible (connected or disconnected) diagrams that involve $m$ particles and $m$ antiparticles. Each such diagram $\mathcal{D}$ can then be decomposed into its connected components. In order to represent all diagrams $\mathcal{D}$ involving $m$ particles and $m$ antiparticles, one needs to sum over the integer partitions of $m$,
\begin{equation}
    m=m_1+2m_2+3m_3\dots\,,
\end{equation}
to establish how many particles and antiparticles will be involved in each of the connected diagrams in $\mathcal{D}$. A particular partition $(m_1,m_2,m_3\dots)$ is associated with the case where the disconnected diagram $\mathcal{D}$ consists of $m_1$ connected diagrams involving 1 particle and 1 antiparticle, $m_2$ connected diagrams involving 2 particles and 2 antiparticles, and so on.
Multiplying these contributions together, we obtain the contribution from the disconnected diagram $\mathcal{D}$
\begin{equation}
    \mathcal{D}= \frac{1}{m_1!} 
    D_{1|1}(\kappa)^{m_1}\times \frac{1}{m_2!} D_{2|2}(\kappa)^{m_2}\times\dots\,.
\end{equation}
Here, we have also multiplied by the appropriate symmetry factors to account for the indistinguishability under permutations of the $m_n$ connected diagrams consisting of $n$ particle-antiparticle pairs for each $n$. Combining these contributions and summing over $m$, the total number of particle-antiparticle pairs, we find the second presentation of the partition function,
\begin{equation}
    \mathcal{Z}(\kappa)=\sum_{m=0}^\infty\sum_{\substack{m_1,m_2,m_3,\dots\geq0\\m_1+2m_2+3m_3+\dots=m}}\prod_{n=1}^m\frac{D_{n|n}(\kappa)^{m_n}}{m_n!}\,.
\end{equation}
Of course, this matches the result in (\ref{eqn:partition_function_as_sum_over_connected}), but the expressions are organized differently. It is this second organization that matches the expansion (\ref{eqn:partition_function_3_1}). Concretely, we claim that
\begin{equation}
    \mathcal{G}^{(m)}(\kappa)=\sum_{\substack{m_1,m_2,m_3,\dots\geq0\\m_1+2m_2+3m_3+\dots=m}}\prod_{n=1}^m\frac{D_{n|n}(\kappa)^{m_n}}{m_n!}\,.\label{eqn:mth_term_interpretation}
\end{equation}
In particular, for the first few values of $m$, we have,
\begin{align}
    \mathcal{G}^{(1)}(\kappa)&=D_{1|1}(\kappa)\,,\\
    \mathcal{G}^{(2)}(\kappa)&=D_{2|2}(\kappa)+\frac{1}{2!}D_{1|1}(\kappa)^2\,,\\
    \mathcal{G}^{(3)}(\kappa)&=D_{3|3}(\kappa)+D_{1|1}(\kappa)D_{2|2}(\kappa)+\frac{1}{3!}D_{1|1}(\kappa)^3\,,\\
    \mathcal{G}^{(4)}(\kappa)&=D_{4|4}(\kappa)+D_{1|1}(\kappa)D_{3|3}(\kappa)+\frac{1}{2!}D_{2|2}(\kappa)^2+\frac{1}{2!}D_{1|1}(\kappa)^2D_{2|2}(\kappa)+\frac{1}{4!}D_{1|1}(\kappa)^4\,.
\end{align}
We will now explicitly demonstrate that the relation (\ref{eqn:mth_term_interpretation}) holds and thereby determine $D_{n|n}(\kappa)$.\\\\
First, we use the symmetry of the product of exponentials in (\ref{eqn:Gm_same_epsilon}) to permute the labels of the integration variables $y_j$ without affecting the result. In particular, we can perform the relabeling $y_j\rightarrow y_{p^{-1}(j)}$, and we can rewrite the sum over $q\in S_m$ as a sum over $q'=p^{-1}\circ q\in S_m$. The sum over $p$ is now trivial and produces a factor of $m!$, such that,
\begin{multline}
    \mathcal{G}^{(m)}(\kappa)=\frac{1}{m!}\sum_{q'\,\in\,S_m}\,\text{sgn}(q')\prod_{j=1}^m\int_{-\infty}^{\infty}\frac{\dd x_j}{2\pi}\,\prod_{j=1}^m\int_{-\infty}^{\infty}\frac{\dd y_j}{2\pi} \\
    \times\prod_{j=1}^m\frac{1}{\left(x_j-y_j+i\epsilon\right)\left(x_j-y_{q'(j)}+i\epsilon\right)}\times\prod_{j=1}^m\exp\left[\kappa\; v^+\left(x_j+\frac{i\epsilon}{2}\right)+\kappa\; v^-\left(y_j-\frac{i\epsilon}{2}\right)\right]\,.
\end{multline}
For a fixed permutation $q'$, the integrand can now be written as a product over the independent cycles $c$ of the permutation $q'$:
\begin{equation}
    \mathcal{G}^{(m)}(\kappa)=\frac{1}{m!}\sum_{q'\,\in\,S_m}\prod_{\text{cycles }c}I_{|c|}(\kappa)\,,\label{eqn:G_as_sum_over_cycles}
\end{equation}
where
\begin{multline}
    I_{|c|}(\kappa)=\text{sgn}(c)\prod_{j=1}^{|c|}\int_{-\infty}^{\infty}\frac{\dd x_j}{2\pi}\,\prod_{j=1}^{|c|}\int_{-\infty}^{\infty}\frac{\dd y_j}{2\pi}\,\prod_{j=1}^{|c|}\frac{1}{\left(x_j-y_j+i\epsilon\right)\left(x_j-y_{j+1}+i\epsilon\right)}\\
    \times\prod_{j=1}^{|c|}\exp\left[\kappa\; v^+\left(x_j+\frac{i\epsilon}{2}\right)+\kappa\; v^-\left(y_j-\frac{i\epsilon}{2}\right)\right]\,.
\end{multline}
Here, we are taking $|c|$ to denote the length of the cycle $c$ in the permutation $q'$, and we have relabeled the integration variables for convenience. Additionally, we are taking $y_{|c|+1}$ to mean $y_1$ in this expression.\\\\
Finally, we can express the sum over permutations $q'\in S_m$ in (\ref{eqn:G_as_sum_over_cycles}) as a sum over choices of cycles. Suppose we want to construct a particular permutation $q'\in S_m$. We begin by choosing how many cycles of each possible length the permutation will have. Let this permutation $q'$ have $m_1$ cycles of length $1$, $m_2$ cycles of length $2$, and so on, such that
\begin{equation}
    \sum_{n=1}^m n\,m_n=m\,.
\end{equation}
Any two permutations that have the same corresponding numbers $(m_1,m_2,\dots,m_m)$ will give the same contribution in (\ref{eqn:G_as_sum_over_cycles}). We can build the cycles as follows: first, arrange the particle labels in some order. There are $m!$ choices of how to do so. Assign the first $m_1$ particles to the cycles of length 1 (the first particle to the first cycle, the second particle to the second cycle, and so on), then assign the next $2m_2$ particles to the cycles of length 2, and so on. In this way, we would construct all permutations $q'$, but we would be overcounting since the permutation derived from $q'$ by exchanging two cycles of the same length would be identical to $q'$. To account for this, we need to divide by the product of symmetry factors $1/m_n!$. We would also be overcounting since the permutation derived from $q'$ by cyclically permuting the elements of one of the disjoint cycles in $q'$ would result in a permutation that is identical to $q'$. Thus, for each cycle in the permutation, we need to divide by its length.\\\\
Thus, we can express (\ref{eqn:G_as_sum_over_cycles}) as
\begin{align}
    \mathcal{G}^{(m)}(\kappa)&=\sum_{\substack{m_1,m_2,m_3,\dots\geq0\\m_1+2m_2+3m_3+\dots=m}}\prod_{n=1}^m\frac{D_{n|n}(\kappa)^{m_n}}{m_n!}\,,\label{eqn:disconnected_from_connected}
\end{align}
where
\begin{multline}
    D_{n|n}(\kappa) \equiv \frac{1}{n} I_{n}(\kappa) 
    =-\frac{1}{n}\prod_{j=1}^n\int_{-\infty}^{\infty}\frac{\dd x_j}{2\pi i}\,\prod_{j=1}^n\int_{-\infty}^{\infty}\frac{\dd y_j}{2\pi i} \prod_{j=1}^n\frac{1}{\left(x_j-y_j+i\epsilon\right)\left(x_j-y_{j+1}+i\epsilon\right)}\\
    \times\prod_{j=1}^n\exp\left[\kappa\; v^+\left(x_j+\frac{i\epsilon}{2}\right)+\kappa\; v^-\left(y_j-\frac{i\epsilon}{2}\right)\right]\,,
    \label{eqn:def_connected_diagram_as_cycle}
\end{multline}
is the contribution from the connected diagram associated with $n$ particles and $n$ antiparticles.\\\\
This concludes the proof that the terms $\mathcal{G}^{(m)}(\kappa)$ and $D_{n|n}(\kappa)$ satisfy the combinatorial relations expected of disconnected and connected Feynman diagrams. However, we recognize that the expression (\ref{eqn:def_connected_diagram_as_cycle}) we have found for the connected diagram $D_{n|n}(\kappa)$ in terms of position space integrals is rather opaque. The momentum space Feynman diagrams are often easier to work with and more intuitive. We will examine them in the following subsection.

\subsection{The momentum space representation of the expansion terms}\label{sec:3.3}

The relation (\ref{eqn:disconnected_from_connected}) implies that we can focus on the connected diagrams $D_{n|n}(\kappa)$ given in (\ref{eqn:def_connected_diagram_as_cycle}) as the fundamental building blocks of the partition function of the minimal superstring theory. To convert to momentum space, we will evaluate the Fourier transforms of the various factors in the integrand of (\ref{eqn:def_connected_diagram_as_cycle}) and make use of the convolution theorem.
Our conventions for the Fourier transform and its inverse are as follows:
\begin{align}
    \mathcal{F}[f(x,y)](p,q)&=\int_{-\infty}^{+\infty} \frac{\dd x}{2\pi} \int_{-\infty}^{+\infty} \frac{\dd y}{2\pi}\;{f(x,y)\;e^{-i(px+qy)}}\,,\\
    \mathcal{F}^{-1}[F(p,q)](x,y)&=\int_{-\infty}^{+\infty} \dd p\int_{-\infty}^{+\infty} \dd q\;{F(p,q)\;e^{i(px+qy)}}\,.
\end{align}
We begin by evaluating the Fourier transform of $\frac{1}{x-y+i\epsilon}$. Since
\begin{equation}
    \int_{-\infty}^{+\infty}\frac{\dd x}{2\pi}\;\frac{1}{x-y+i\epsilon}e^{-ipx}=-i\,e^{-p\epsilon-ipy}\,\theta(p)\,,
\end{equation}
where $\theta(p)$ denotes the Heaviside step function, and
\begin{equation}
    \int_{-\infty}^{+\infty}\frac{\dd y}{2\pi}\;e^{-p\epsilon-ipy}e^{-iqy}=e^{-p\epsilon}\,\delta(p+q)\,,
\end{equation}
where $\delta(p+q)$ denotes the Dirac delta function, we find
\begin{equation}
    \mathcal{F}\left[\frac{1}{x-y+i\epsilon}\right](p,q)=-i\,e^{-p\epsilon}\theta(p)\,\delta(p+q)\,.\label{eqn:Fourier_single_denominator}
\end{equation}
We can treat (\ref{eqn:def_connected_diagram_as_cycle}) as a Fourier transform evaluated at external momenta $(\vec{p},\vec{q})=(0,0)$,
\begin{align}
    D_{n|n}(\kappa)&=\frac{(-1)^{n+1}}{n}\mathcal{F}\Bigg[\prod_{j=1}^n\frac{1}{x_j- y_j+i\epsilon}\prod_{j=1}^n\frac{1}{x_j-y_{j+1}+i\epsilon}\times\nonumber\\
    &\times\prod_{j=1}^n\exp\left[\kappa\,v^+\left(x_j+\frac{i\epsilon}{2}\right)\right]\prod_{j=1}^n\exp\left[\kappa\,v^-\left(y_j-\frac{i\epsilon}{2}\right)\right]\Bigg](0,0)\,.\label{eqn:Dnn_as_Fourier}
\end{align}
In the following, we will denote by $g_\pm(p)$ the corresponding Fourier transforms of the exponential factors,
\begin{equation}
    g_{\pm}(p)\equiv\mathcal{F}\left[\exp\left[\kappa v^\pm\left(z\pm\frac{i\epsilon}{2}\right)\right]\right](p)\,.
\end{equation}
First, we use the convolution theorem to evaluate the Fourier transform of the last three factors in (\ref{eqn:Dnn_as_Fourier}). 
Using (\ref{eqn:Fourier_single_denominator}), it follows that\footnote{Here, we define $p_0'\equiv p_n'$.}
\begin{multline}
    \mathcal{F}\Bigg[\prod_{j=1}^n\frac{1}{x_j-y_{j+1}+i\epsilon}\prod_{j=1}^n\exp\left[\kappa\,v^+\left(x_j+\frac{i\epsilon}{2}\right)\right]\prod_{j=1}^n\exp\left[\kappa\,v^-\left(y_j-\frac{i\epsilon}{2}\right)\right]\Bigg](p_j,q_j)\\
    =(-i)^n\prod_{j=1}^n\int_0^\infty\dd p_j'\,\exp\left(-\epsilon\sum_{j=1}^n p_j'\right)\prod_{j=1}^ng_+(p_j-p_j')\prod_{j=1}^ng_-(q_j+p_{j-1}')\,.
\end{multline}
Using the convolution theorem again, we have
\begin{multline}
    D_{n|n}(\kappa)=-\frac{1}{n}\prod_{j=1}^n\int_0^\infty\dd p_j\,\prod_{j=1}^n\int_0^\infty\dd  p_j'\,\exp\left(-\epsilon\sum_{j=1}^n\left(p_j+p_j'\right)\right)\\
    \times\prod_{j=1}^n g_+(-p_j-p_j')\prod_{j=1}^ng_-(p_j+p_{j-1}')\,.
\end{multline}
Note that for minimal superstring theory, the potential satisfies $v^+(-z)=v^-(z)$, so $g_+(-p)=g_-(p)$. Thus, we find,
\begin{equation}
    D_{n|n}(\kappa)=-\frac{1}{n}\prod_{j=1}^n\int_0^\infty\dd p_j\,\prod_{j=1}^n\int_0^\infty\dd p_j'\,\exp\left(-\epsilon\sum_{j=1}^n\left(p_j+p_j'\right)\right)\prod_{j=1}^ng_-(p_{j-1}'+p_j)\,g_-(p_j+p_j')\,.
\end{equation}
Notice that the expression has cyclic symmetry in the $2n$ momenta, and we can perform a relabeling to obtain the final result,\footnote{Here, we define $p_{2n+1} \equiv p_1$.}
\begin{equation}
    D_{n|n}(\kappa)=-\frac{1}{n}\prod_{j=1}^{2n}\int_0^\infty\dd p_j\,\exp\left(-\epsilon\sum_{j=1}^{2n}p_j\right)\prod_{j=1}^{2n}g_-(p_j+p_{j+1})\,.
\end{equation}
For $n=1$, this takes the form,
\begin{equation}
\begin{split}
    D_{1|1}(\kappa)&=-\int_0^\infty\dd p_1\,\int_0^\infty\dd p_2\,e^{-\epsilon(p_1+p_2)}g_-(p_1+p_2)^2\\
    &=-\int_0^\infty\dd P \,P \,e^{-\epsilon P}g_-(P)^2\,.
    \label{eqn:D11_in_momentum_space_final}
\end{split}
\end{equation}
Indeed, one can evaluate these integrals numerically. Numerical results for $D_{1|1}(\kappa)$ are shown in Figure \ref{fig:D11} for various integers $k$.\\\\
In appendix \ref{app:pure_sugra}, we will use the momentum space expressions derived in this section to study the special case of the $(2,4)$ minimal superstring theory i.e., pure supergravity.

\begin{figure}
    \centering
    \includegraphics[width=0.7\textwidth]{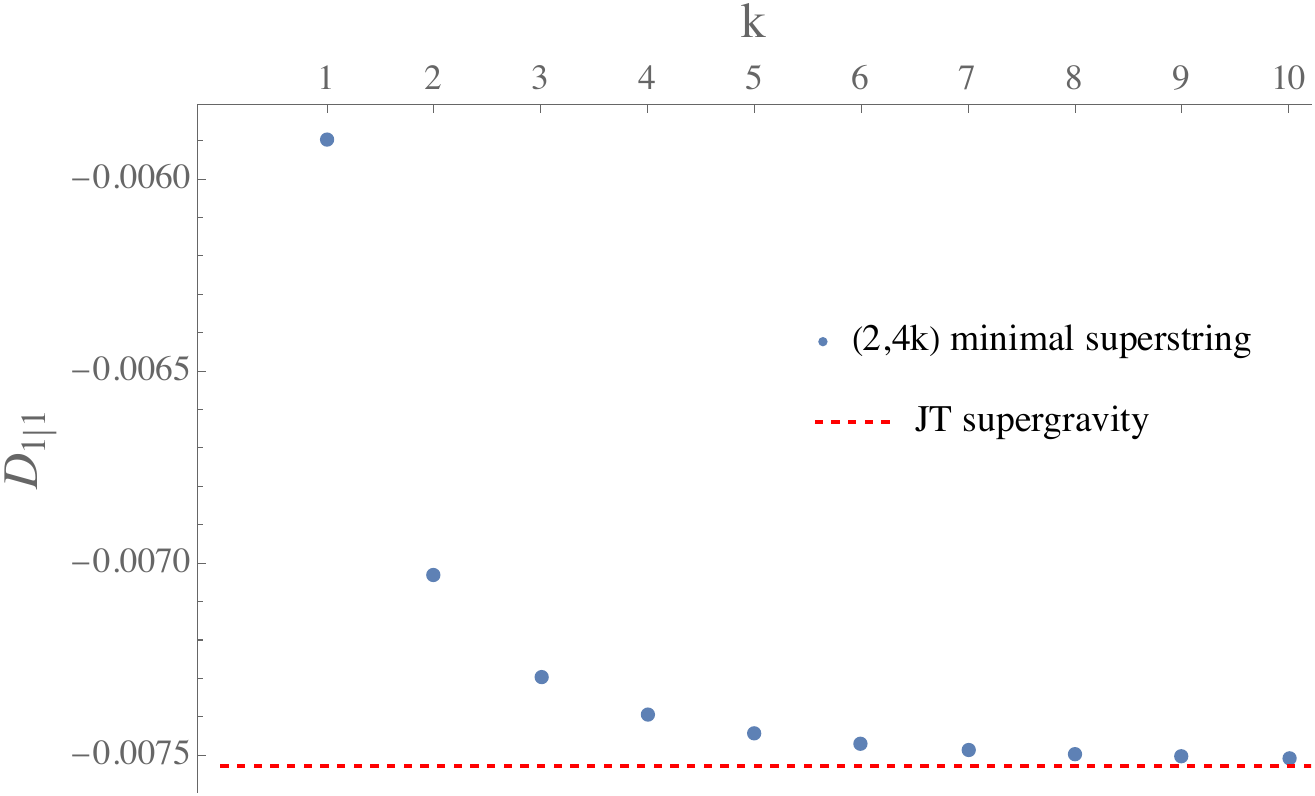}
    \caption{Numerical results for $D_{1|1}(\kappa)$, the connected diagram involving one particle and one antiparticle, evaluated at $\kappa=0.1$. The $(2,4k)$ minimal superstring theory results as well as the JT supergravity limit are shown.}
    \label{fig:D11}
\end{figure}
\section{The Hilbert space formulation}
\label{sec:4}

We have seen how the complete non-perturbative partition function of the $(2,4k)$ minimal superstring theory arises from the double-scaling limit of the Fredholm determinant expansion of a unitary matrix integral. In this section, we will express the partition function directly as the Fredholm determinant of an operator defined on a particular Hilbert space which we will refer to as the quantum gravity Hilbert space. We then show that our results can be precisely expressed in the language of a one-dimensional theory of free Weyl fermions.\footnote{While not explicitly written in this language, the proof of the Borodin-Okounkov theorem \cite{borodin2000fredholm} which applies in the finite-$N$ case also makes use of the formalism of free fermions and bosonization. See Ref.~\cite{Murthy:2022ien} for details. The major difference between the finite-$N$ case and the superstring case is that in the finite-$N$ case, the Hilbert space is naturally identified with $\ell^2(\mathbb{N})$, with orthonormal basis states $\ket{n}$ naturally labeled by non-negative integers $n\in\mathbb{N}$, whereas in the superstring case, the Hilbert space is naturally identified with $L^2(\mathbb{R})$. In the case of the latter, one often prefers to work with the associated rigged Hilbert space (momentum space) where the natural basis is the continuous family of states $\ket{p}$ labeled by $p\in\mathbb{R}$.}

\subsection{Double-scaling at the operator level}

In the finite-$N$ case, the Fredholm determinant expansion is given in (\ref{eqn:Fredholm_det_expansion}) which we reproduce here for convenience,
\begin{equation}
    \frac{Z(N,g,t_l^\pm)}{Z^{(0)}(N,g,t_l^\pm)}=1+\sum_{m=1}^\infty G_N^{(m)}\,.
    \label{eqn:Fred_det_expansion_dsatol}
\end{equation}
As shown in Appendix \ref{app:FredProof}, the terms in this expansion take the form,
\begin{equation}
    G_N^{(m)}=\frac{(-1)^m}{m!}\sum_{0\leq n_1,n_2,\dots,n_m}\det[f_N(n_i,n_j)]_{i,j=1,\dots,m}\,,
\label{eqn:Fred_app_dsatol}
\end{equation}
where
\begin{multline}
    f_N(n_i,n_j)=\oint_{r_-}\frac{\dd u}{2\pi i}\oint_{r_+}\frac{\dd v}{2\pi i}u^{n_i}v^{-n_j}\left(\frac{u}{v}\right)^{N+1}\frac{1}{1-u/v} \\
    \times\exp\left[\frac{N}{g}\sum_{l=1}^k\left(\frac{t_l^-}{l}\left(u^l-v^l\right)-\frac{t_l^+}{l}\left(u^{-l}-v^{-l}\right)\right)\right]\,.
\label{eqn:def_coeffs}
\end{multline}
The right side of (\ref{eqn:Fred_app_dsatol}) is a Fredholm determinant associated with a particular operator $T_N$ on $\ell^2(\mathbb{N})$, the space of all square-summable complex sequences. The operator in question can be defined through its action on the orthonormal basis $\{\ket{n}\}_{n\in\mathbb{N}}$,
\begin{equation}
    T_N\ket{n}=\sum_{n'=0}^\infty f_N(n,n')\ket{n'}\,.
\end{equation}
The Fredholm determinant associated with $T_N$ is
\begin{equation}
    \det(\mathbb{1}-\mu\;T_N)=1+\sum_{m=1}^\infty(-\mu)^m\,\Tr\,\Lambda^m(T_N)\,,
\end{equation}
where $\mu$ is a complex parameter, and 
\begin{equation}
    \Tr\,\Lambda^m(T_N)=\frac{1}{m!}\sum_{0\leq n_1,\dots,n_m}\det[f_N(n_i,n_j)]_{i,j=1,\dots,m}\,.
\label{eqn:TrLambda}
\end{equation}
This means that for finite $N$, the right side of (\ref{eqn:Fred_det_expansion_dsatol}) is simply the Fredholm determinant $\det(\mathbb{1}-\mu T_N)$ evaluated at $\mu=1$.\\\\
We have already shown that each term in the expansion (\ref{eqn:Fred_det_expansion_dsatol}) admits a double-scaling limit. The remaining question we need to focus on is how to describe the double-scaling limit for the operator $T_N$. More precisely, we need to understand how the states $\ket{n}$ and the coefficients $f_N(n_i,n_j)$ behave under double scaling. Starting with the definition (\ref{eqn:def_coeffs}), we can make the substitutions,
\begin{align}
    u=\left(-1+\frac{i\varepsilon\lambda}{k}\right)^{-1}\,,\qquad v=\left(-1+\frac{i\varepsilon\sigma}{k}\right)^{-1}\,,\qquad n_i=\frac{k}{\varepsilon}p_i\,,\qquad n_j=\frac{k}{\varepsilon}p_j\,.
\end{align}
We are interested in taking the limit $N\to\infty , \, \varepsilon\to 0$ while keeping fixed the quantity
\begin{equation}
    \kappa=\frac{1}{2k}\binom{2k}{k}^{-1}N\varepsilon^{2k+1}\,.
\end{equation}
We have already shown that, in this limit,
\begin{equation}
    \left(\frac{u}{v}\right)^N\exp\left[\frac{N}{g}\sum_{l=1}^k\left(\frac{t_l^-}{l}\left(u^l-v^l\right)-\frac{t_l^+}{l}\left(u^{-l}-v^{-l}\right)\right)\right]\rightarrow\exp\left[\kappa\left(v^+(\lambda_i)+v^-(\sigma_i)\right)\right]\,,
\end{equation}
where $v^{\pm}(\xi)$ are the double-scaled effective potentials given in (\ref{eqn:veff_pm}).
Additionally, the factor $\frac{u/v}{1-u/v}$ becomes
\begin{equation}
    \frac{u/v}{1-u/v}\rightarrow\frac{ik/\varepsilon}{\lambda-\sigma}\,.
\end{equation}
As earlier, the integration contours change from circles of radii $r_-<1<r_+$ to lines above and below the real axis,
\begin{equation}
    \oint_{r_-}\frac{\dd u}{2\pi i}\rightarrow-\frac{\varepsilon}{2\pi k}\int_{\mathcal{C}_+}\dd\lambda\,,
    \qquad\qquad\oint_{r_+}\frac{\dd v}{2\pi i}\rightarrow-\frac{\varepsilon}{2\pi k}\int_{\mathcal{C}_-}\dd\sigma\,.
\end{equation}
The only remaining terms become
\begin{align}
    u^{n_i}&=(-1)^{n_i}\left(1-\frac{i\varepsilon\lambda}{k}\right)^{-kp_i/\varepsilon}\rightarrow e^{ip_i\lambda}\,,\label{eqn:u_n_i}\\
    v^{-n_j}&=(-1)^{-n_j}\left(1-\frac{i\varepsilon\sigma}{k}\right)^{kp_j/\varepsilon}\rightarrow e^{-ip_j\sigma}\,.\label{eqn:v_n_j}
\end{align}
In the finite-$N$ case, in the sums appearing in (\ref{eqn:Fred_app_dsatol}) and (\ref{eqn:TrLambda}), the variables $n_i$ range over the non-negative integers. Since $p_i$ is proportional to $\varepsilon n_i$, in the small $\varepsilon\rightarrow0$ limit of double scaling, the variables $p_i$ will range over the non-negative reals, and the sums in (\ref{eqn:Fred_app_dsatol}) and (\ref{eqn:TrLambda}) should naturally be replaced with integrals over $[0,\infty)$. The coefficients $f_N(n_i,n_j)$ which define the trace-class operator $T_N$ over $\ell^2(\mathbb{N})$ therefore become a kernel $f(p_i,p_j)$ which defines a trace-class operator $T$ naturally expressed in the momentum space basis associated with the Hilbert space $L^2(\mathbb{R})$:
\begin{equation}
    T\ket{p}=\int_{-\infty}^{+\infty}\dd p'\;f(p,p')\ket{p'}\,.
\end{equation}
The factors of $(-1)^{n_i}$ and $(-1)^{n_j}$ appearing in (\ref{eqn:u_n_i}) and (\ref{eqn:v_n_j}) can be discarded since in all of the final expressions for physical quantities such as (\ref{eqn:TrLambda}), the trace on the left side ensures the cancellation of all of these terms.\footnote{One can see this explicitly for (\ref{eqn:TrLambda}) since the determinant on the right side can be expressed as a sum over permutations which can further be expressed as sums over cycles. For each $f(p_i,p_j)$ appearing in such a cycle, there will be an associated $f(p_j,p_k)$ appearing in the same cycle which ensures that each $p_j$ either never appears, or appears exactly twice in a cycle, leading to two copies of the factor $(-1)^{n_j}$.}
Finally, we need to include a factor of $k/\varepsilon$ that arises from changing the sum over $n$ to an integral over $p$.\\\\
We find that the kernel $f(p,p')$ which was originally defined for $p,p'>0$ can be used to define the trace-class operator $T$, with momentum space representation
\begin{equation}
    f(p,p')=\braket{p'|T|p}=\theta(p)\theta(p')\int_{\mathcal{C}_+}\frac{\dd\lambda}{2\pi}\int_{\mathcal{C}_-}\frac{\dd\sigma}{2\pi}\,e^{ip\lambda}e^{-ip'\sigma}\frac{i}{\lambda-\sigma}\exp\left[\kappa(v^+(\lambda)+v^-(\sigma))\right]\,.
\end{equation}
Next, we show that the operator $T$ is self-adjoint. The adjoint of $T$ is the operator $T^\dagger$ on $L^2(\mathbb{R})$ which can be expressed in the momentum space basis as,
\begin{equation}
    T^\dagger\ket{p}=\int_{-\infty}^{+\infty}\dd p'\;f(p',p)^*\ket{p'}\,.
\end{equation}
Since the effective potentials $v^\pm$ satisfy $(v^+(s))^*=v^-(s^*)$, it follows that
\begin{equation}
    f(p',p)^*=f(p,p')\,,
\end{equation}
and therefore that $T^\dagger=T$. Thus, the trace-norm of the operator $T$ is
\begin{equation}
    \|T\|_1=\Tr\left(\sqrt{T^\dagger T}\right)=\Tr(T)=\int_{-\infty}^{+\infty}\dd p\;\braket{p|T|p}=\int_{-\infty}^{+\infty}\dd p\;f(p,p)\,.
\end{equation}
Performing the integral over the momentum $p$ and noting that $\text{Im}[\lambda-\sigma]>0$, we find,
\begin{equation}
    \|T\|_1=\Tr(T)=-\int_{\mathcal{C}_+}\frac{\dd\lambda}{2\pi}\int_{\mathcal{C}_-}\frac{\dd\sigma}{2\pi}\frac{1}{(\lambda-\sigma)^2}\exp[\kappa(v^+(\lambda)+v^-(\sigma))]=-\mathcal{G}^{(1)}(\kappa)\,.
\end{equation}
Thus, the operator $T$ is trace-class if and only if the preceding expression converges. We have shown that this is indeed the case in section \ref{sec:3.1}, so we conclude that $T$ is trace-class. This allows us to construct its associated Fredholm determinant and to show that it exactly matches the expression (\ref{eqn:BOGC_expansion_string}) we have found for the partition function of the $(2,4k)$ minimal superstring theory.\footnote{One can show that the Fredholm determinant associated with any trace-class operator is well-defined and convergent. See \cite{lax2014functional} for details.} This concludes the proof that expansion for the partition function in \eqref{eqn:Fredholm_det_expansion} is indeed convergent.

\subsection{The partition function as a Fredholm determinant}

We will now verify that the terms $\mathcal{G}^{(m)}(\kappa)$ in the expansion of the partition function of the $(2,4k)$ minimal superstring theory are reproduced by the terms appearing in the expression for the Fredholm determinant associated with the operator $T$,
\begin{align}
    \mathcal{G}^{(m)}(\kappa)&=(-1)^m\;\Tr\,\Lambda^m(T)\label{eqn:mathcalG_asFredholm}\\
    &=\frac{(-1)^m}{m!}\int_{-\infty}^{+\infty}\dd p_1\int_{-\infty}^{+\infty}\dd p_2\dots\int_{-\infty}^{+\infty}\dd p_m\;\det[f(p_i,p_j)]_{i,j=1,\dots,m}\,.
\end{align}
For $m=1$, we have already calculated the right side of the previous expression in the preceding subsection,
\begin{align}
    -\Tr(T)=\int_{\mathcal{C}_+}\frac{\dd\lambda}{2\pi}\int_{\mathcal{C}_-}\frac{\dd\sigma}{2\pi}\,\frac{1}{(\lambda-\sigma)^2}\exp\left[\kappa(v^+(\lambda)+v^-(\sigma))\right]\,.
\end{align}
Thus, we indeed recover the expression (\ref{eqn:mathcalGmkappa}) for $\mathcal{G}^{(1)}(\kappa)$.\\\\
For $m>1$, we have
\begin{equation}
\label{eqn:LambdaT_expansion}
\begin{split}
    (-1)^m\,\Tr\,\Lambda^m(T)&=\frac{(-1)^m}{m!}\sum_{q\in S_m}\text{sgn}(q) \prod_{i=1}^{m} \int_{-\infty}^{+\infty}\dd p_i \, \prod_{j=1}^mf(p_j,p_{q(j)})\\
    &=\frac{1}{m!}\sum_{q\in S_m}\text{sgn}(q)\prod_{j=1}^m\int_{\mathcal{C}_+}\frac{\dd\lambda_j}{2\pi}\prod_{j=1}^m\int_{\mathcal{C}_-}\frac{\dd\sigma_j}{2\pi}\prod_{i=1}^m\frac{1}{\lambda_i-\sigma_i}\prod_{i=1}^m\frac{1}{\lambda_i-\sigma_{q^{-1}(i)}}\\
    &\qquad\qquad\qquad\qquad\qquad\qquad\qquad\qquad\qquad\times\prod_{j=1}^m\exp[\kappa(v^+(\lambda_j)+v^-(\sigma_j))]\\
    &=\frac{1}{m!}\prod_{j=1}^m\int_{\mathcal{C}_+}\frac{\dd\lambda_j}{2\pi}\prod_{j=1}^m\int_{\mathcal{C}_-}\frac{\dd\sigma_j}{2\pi}\prod_{i=1}^m\frac{1}{\lambda_i-\sigma_i}\det\left[\frac{1}{\lambda_i-\sigma_j}\right]_{i,j=1,\dots,m}\\
    &\qquad\qquad\qquad\qquad\qquad\qquad\qquad\qquad\qquad\times\prod_{j=1}^m\exp[\kappa(v^+(\lambda_j)+v^-(\sigma_j))]\,.
\end{split}
\end{equation}
We can use the argument of Refs.~\cite{tracy2013diagonal, tracy2017natural} to perform the replacement
\begin{equation}
    \prod_{i=1}^m\frac{1}{\lambda_i-\sigma_i}\det\left[\frac{1}{\lambda_i-\sigma_j}\right]_{i,j=1,\dots,m}\rightarrow\frac{1}{m!}\det\left[\frac{1}{\lambda_i-\sigma_j}\right]^2_{i,j=1,\dots,m}\,,
\end{equation}
and use the Cauchy determinant identity,
\begin{equation}
    \det\left[\frac{1}{\lambda_i-\sigma_j}\right]_{i,j=1,\dots,m}=\frac{\prod_{1\leq j<i\leq m}(\lambda_i-\lambda_j)(\sigma_j-\sigma_i)}{\prod_{i,j=1}^m(\lambda_i-\sigma_j)}\,,
\end{equation}
to recover the expression for $\mathcal{G}^{(m)}(\kappa)$ in (\ref{eqn:mathcalGmkappa}). Thus, we conclude that (\ref{eqn:mathcalG_asFredholm}) is indeed true and that the partition function of the $(2,4k)$ minimal superstring theory is equal to the Fredholm determinant associated with the operator $T$,
\begin{equation}
    \mathcal{Z}(\kappa)=\det(\mathbb{1}-T)\,.
\end{equation}
One can also immediately recover an expression for the free energy of the theory in terms of the operator $T$ by using the following property of the Fredholm determinant,
\begin{equation}
    F(\kappa)=\log(\mathcal{Z}(\kappa))=\log \det(\mathbb{1}-T)=\Tr\log(\mathbb{1}-T)=-\sum_{n=1}^\infty\frac{1}{n}\Tr(T^n)\,.
\end{equation}
Thus, the connected diagrams $D_{n|n}(\kappa)$ can be expressed as,
\begin{multline}
    D_{n|n}(\kappa)=-\frac{1}{n}\Tr(T^n)\\
    =-\frac{1}{n}\prod_{j=1}^n\int_{\mathcal{C}_+}\frac{\dd\lambda_j}{2\pi i}\prod_{j=1}^n\int_{\mathcal{C}_-}\frac{\dd\sigma_j}{2\pi i}\prod_{j=1}^n\left(\frac{1}{(\lambda_j-\sigma_j)(\lambda_j-\sigma_{j-1})}\exp[\kappa(v^+(\lambda_j)+v^-(\sigma_j))]\right)\,.
\end{multline}
After a simple relabeling, this matches the result in (\ref{eqn:def_connected_diagram_as_cycle}).

\subsection{The minimal superstring as a theory of free fermions}

Our goal in this subsection is to make precise the claim that the complete non-perturbative description of the ungapped phase of type 0B $(2,4k)$ minimal superstring theory can be reproduced as that of a target space theory of free one-dimensional Weyl fermions living on the real line. The present discussion closely resembles that of Ref.~\cite{Murthy:2022ien}, which we appropriately modify to account for $N$ being infinite in our case.\\\\
We begin by defining the algebra of free Weyl fermions in momentum space. The algebra is generated by two types of operators, $\psi(p)$ and $\bar{\psi}(p)$, which satisfy the following anti-commutation relations,
\begin{equation}
    \{\psi(p),\psi(q)\}=0\,,\qquad\{\bar{\psi}(p),\bar{\psi}(q)\}=0\,,\qquad\{\psi(p),\bar{\psi}(q)\}=\delta(p+q)\,.
\end{equation}
We define the adjoint on the fermionic algebra by its action on the generators,
\begin{equation}
    \psi(p)^\dagger=\bar{\psi}(-p)\,,\qquad\bar{\psi}(p)^\dagger=\psi(-p)\,.
\end{equation}
A representation of this algebra is obtained by considering the action of the algebra on a reference state $\ket{0}$ which we will refer to as the Dirac vacuum. We choose $\ket{0}$ to satisfy,
\begin{equation}
    \psi(p)\ket{0}=\bar{\psi}(p)\ket{0}=0\,,\quad\text{for } p>0\,.
\end{equation}
The states of the rigged Hilbert space associated with the one-particle Hilbert space are spanned by the collection $\psi(p)\ket{0}$ and $\bar{\psi}(p)\ket{0}$ for $p<0$. The full space of states of the fermionic theory is the fermionic Fock space built out of the one-particle Hilbert space.\footnote{We emphasize that, although we are working in momentum space, one should not confuse the one-particle Hilbert space, which is simply $L^2(\mathbb{R})$ and admits a countable orthonormal basis, with the set of momentum space basis kets $\psi(p)\ket{0}$, which are not elements of the Hilbert space, but instead are elements of the associated rigged Hilbert space. The present discussion is no different than the usual discussion of momentum space states in the case of a particle on a line in textbook quantum mechanics.}\\\\
In order to reproduce the expressions found in the case of the minimal superstring, the fermions need to be dressed by appropriate operators in the Heisenberg picture:
\begin{equation}
    \Psi(p)=U\psi(p)U^{-1}\,,\qquad\bar{\Psi}(p)=U\bar{\psi}(p)U^{-1}\,.
\end{equation}
Despite the dressing, the theory remains one of free fermions. This can easily be seen by performing a change of basis where the dressing is absorbed into the definition of the vacuum,
\begin{equation}
    \ket{\Omega}=U^{-1}\ket{0}\,.
\end{equation}

\subsubsection{The dressed fermions}

Let us present the construction of the operators $U$ which we will use to define the theory. In the following, we will also utilize the position space fermionic operators,\footnote{We are overloading the notation $\psi$ for convenience. One can deduce that we are referring to a momentum space fermion when the argument is a momentum variable such as $p$ or $q$ and that we are referring to a position space fermion when the argument is a position variable such as $x$, $z$, $\lambda$, or $\sigma$.}
\begin{equation}
    \psi(z)=\int_{-\infty}^{+\infty}\dd p\,e^{ipz}\psi(p)\,,\qquad\bar{\psi}(z)=\int_{-\infty}^{+\infty}\dd p\,e^{ipz}\bar{\psi}(p)\,.
\end{equation}
We first define $\Tilde{v}^\pm(p)$ to be the Fourier transform of the effective potential $\kappa\,v^\pm(x)$:
\begin{equation}
    \Tilde{v}^\pm(p)=\int_{-\infty}^{+\infty}\frac{\dd x}{2\pi}\,\kappa\,v^\pm(x)\, e^{ipx}\,.
\end{equation}
For the effective potentials in \eqref{eqn:veff_pm}, it follows that $\Tilde{v}^\pm(-p)=\Tilde{v}^\mp(p)$ and $(\Tilde{v}^\pm{}(p))^*=-\Tilde{v}^\mp(p)$. We will also define the following bosonic operators,
\begin{equation}
    a(p)=\int_{-\infty}^{+\infty}\dd q\,\bar{\psi}(p-q)\psi(q)\,,\qquad S_{\pm}=\int_0^\infty\dd p\,\Tilde{v}^\pm(p)a(\pm p)\,.
\end{equation}
One can show that the position space fermion operators satisfy the following commutation relations,
\begin{equation}
    [a(p),\psi(z)]=-e^{-ipz}\psi(z)\,,\qquad[a(p),\bar{\psi}(z)]=e^{-ipz}\bar{\psi}(z)\,.
\end{equation}
Finally, we define the dressing operators
\begin{equation}
    U=\exp(S_+)\exp(S_-)\,,\qquad\qquad U^{-1}=\exp(-S_-)\exp(-S_+)\,.
\end{equation}
It follows that $U$ is unitary since $S_+^\dagger=- S_-$ and $[S_+, S_-]=0$.
One can show that the dressed position space fermions are related to the original position space fermions via\footnote{Note that a similar simple relationship will not exist between the dressed momentum space fermions and the original momentum space fermions. This is the reason why it is more convenient to work with the position space fermions.}
\begin{equation}
    \Psi(z)=e^{\kappa v^-(z)}\psi(z)\,,\qquad\qquad\bar{\Psi}(z)=e^{\kappa v^+(z)}\bar{\psi}(z)\,.
\end{equation}
The dressed momentum space fermions  can be expressed in terms of the dressed position space fermions as 
\begin{equation}
    \Psi(p)=\int_{\mathcal{C}_-}\frac{\dd\sigma}{2\pi}\,e^{-ip\sigma}\,\Psi(\sigma)\,,\qquad\qquad\bar{\Psi}(p)=\int_{\mathcal{C}_+}\frac{\dd\lambda}{2\pi}\,e^{-ip\lambda}\,\bar{\Psi}(\lambda)\,.
\end{equation}

\subsubsection{The partition function as a trace}

We will show that the partition function of the minimal superstring theory can be recast as a trace over the states in the Hilbert space of the theory of dressed free fermions. In defining the physical Hilbert space one has to restrict to the states spanned by $\Psi(p)\ket{0}$, with $p>0$. This restriction can be justified by noting that the state $\bar{\Psi}(p)\ket{0}$, with $p\in\mathbb{R}$, is not normalizable. Moreover, the $p$-integral involving the states spanned by $\Psi(p)\ket{0}$, with $p<0$, diverges at $p \to -\infty$.\footnote{In principle, one could consider states with $p\in[p_0,\infty)$ where $p_0$ is some real constant, but this would simply shift the effective potentials $v^{\pm}(x)$ by $\pm ip_{0}x$.}
We first show that the $m$th term in the Fredholm determinant expansion of the partition function of the minimal superstring theory can be expressed as
\begin{equation}
    \mathcal{G}^{(m)}(\kappa)=\frac{(-1)^m}{m!}\prod_{j=1}^m\int_0^\infty\dd p_j\,\braket{0|\bar{\Psi}(-p_m)\dots\bar{\Psi}(-p_1)\Psi(p_1)\dots\Psi(p_m)|0}\,.\label{eqn:fermion_correlator}
\end{equation}
Since the dressed momentum space fermions can be rewritten as
\begin{align}
    \Psi(p_j)&=\int_{\mathcal{C}_-}\frac{\dd\sigma_j}{2\pi}\;e^{-ip_j\sigma_j}e^{\kappa v^-(\sigma_j)}\int_{-\infty}^{+\infty}\dd q_j\;e^{iq_j\sigma_j}\psi(q_j)\,,\\
    \bar{\Psi}(-p_j)&=\int_{\mathcal{C}_+}\frac{\dd\lambda_j}{2\pi}\;e^{+ip_j\lambda_j}e^{\kappa v^+(\lambda_j)}\int_{-\infty}^{+\infty}\dd \bar{q}_j\;e^{i\bar{q}_j\lambda_j}\bar{\psi}(\bar{q}_j)\,,
\end{align}
the expression on the right of (\ref{eqn:fermion_correlator}) can be evaluated in terms of ordinary free fermion correlators.
It follows by induction that
\begin{equation}
    \braket{0|\bar{\psi}(\bar{q}_m)\dots\bar{\psi}(\bar{q}_1)\psi(q_1)\dots\psi(q_m)|0}=\det\left(\delta(q_i+\bar{q}_j)\right)_{i,j=1}^m\,\prod_{j=1}^m\theta(-q_j)\,.
\end{equation}
Thus, the correlator of dressed fermions takes the form,
\begin{multline}
    \braket{0|\bar{\Psi}(-p_m)\dots\bar{\Psi}(-p_1)\Psi(p_1)\dots\Psi(p_m)|0}\\
    =\prod_{j=1}^m\int_{\mathcal{C}_+}\frac{\dd\lambda_j}{2\pi}\prod_{j=1}^m\int_{\mathcal{C}_-}\frac{\dd\sigma_j}{2\pi}\prod_{j=1}^m e^{ip_j(\lambda_j-\sigma_j)}
    \times\det\left[\frac{1}{i(\sigma_i-\lambda_j)}\right]_{i,j=1,\dots,m}\prod_{j=1}^m e^{\kappa(v^+(\lambda_j)+v^-(\sigma_j))}\,.
\end{multline}
Evaluating the integrals over the momenta on the right side of (\ref{eqn:fermion_correlator}), we find
\begin{multline}
    \frac{(-1)^m}{m!}\prod_{j=1}^m\int_0^\infty\dd p_j\,\braket{0|\bar{\Psi}(-p_m)\dots\bar{\Psi}(-p_1)\Psi(p_1)\dots\Psi(p_m)|0}\\
    =\frac{1}{m!}\prod_{j=1}^m\int_{\mathcal{C}_+}\frac{\dd\lambda_j}{2\pi}\prod_{j=1}^m\int_{\mathcal{C}_-}\frac{\dd\sigma_j}{2\pi}\;\prod_{j=1}^m e^{\kappa(v^+(\lambda_j)+v^-(\sigma_j))}\times\det\left[\frac{1}{\lambda_i-\sigma_j}\right]_{i,j=1,\dots,m} \prod_{j=1}^m\frac{1}{\lambda_j-\sigma_j}\,.
\end{multline}
The right side equals the expression in (\ref{eqn:LambdaT_expansion}) for $(-1)^m\Tr\,\Lambda^m(T)$, which in turn equals $\mathcal{G}^{(m)}(\kappa)$.\\\\
We now provide an interpretation of this identity. Consider the Fock space spanned by dressed fermion modes $\Psi(p)$ with $p>0$. Due to the fermionic nature of the excitations, the modes need to be anti-symmetrized, meaning that for a collection of momenta $(p_1,\dots,p_m)$, we should correct for the overcounting of states when permuting the operators $\Psi(p_i)$ and $\Psi(p_j)$. Thus, when we evaluate traces over the fermionic Fock space spanned by these modes, we will need to introduce a factor of $1/m!$ to account for the anti-symmetry. The trace of an operator $\mathcal{O}$ in this space therefore takes the form,
\begin{equation}
    \Tr(\mathcal{O})=\sum_{m=0}^\infty\frac{1}{m!}\prod_{j=1}^m\int_0^\infty\dd p_j\;\braket{0|\bar{\Psi}(-p_m)\dots\bar{\Psi}(-p_1)\mathcal{O}\Psi(p_1)\dots\Psi(p_m)|0}\,,
\end{equation}
where we have used the fact that the adjoint acts on the dressed fermions analogously to how it acts on the ordinary free fermions,
\begin{equation}
    \Psi(p)^\dagger=\bar{\Psi}(-p)\,,\qquad\bar{\Psi}(p)^\dagger=\Psi(-p)\,.
\end{equation}
Thus, we can interpret the partition function of the $(2,4k)$ minimal superstring as the Witten index of the dressed fermionic theory,
\begin{align}
    \mathcal{Z}=\Tr(-1)^F=\sum_{m=0}^\infty\frac{1}{m!}\prod_{j=1}^m\int_0^\infty\dd p_j\;\braket{0|\bar{\Psi}(-p_m)\dots\bar{\Psi}(-p_1)(-1)^F\Psi(p_1)\dots\Psi(p_m)|0}\,.
\end{align}
\section{Correlators of the eigenvalue density}
\label{sec:5}

The eigenvalue density $\rho(z)$ is a key quantity in the study of matrix integrals. 
In this section, we obtain the exact functional forms of the correlators of this quantity using the Fredholm determinant expansion of the unitary matrix integral. We show that these correlators have well-defined double-scaling limits which we interpret in the dual string theory language in terms of insertions of $\nb$-boundaries.\\\\
To simplify the calculations, we introduce a set of auxiliary quantities called resolvents,
\begin{equation}
    R^\pm(z)=\frac{1}{N}\Tr\left(\left(z\mathbb{1}-U\right)^{-1}\right)\,,\qquad|z|\neq1\,.
\end{equation}
Here, the resolvent with the positive label is originally defined for $|z|>1$, while the resolvent with the negative label is originally defined for $|z|<1$. These quantities differ since the definition of the matrix inverse has different expressions in the two cases,
\begin{align}
    R^+(z)&=\frac{1}{Nz}\Tr\left(\mathbb{1}+\sum_{p=1}^\infty z^{-p}U^p\right)=\frac{1}{z}+\frac{1}{N}\sum_{p=1}^\infty \Tr\left(U^p\right)z^{-p-1}\,,\qquad|z|>1\,,\label{eqn:R+_def}\\
    R^-(z)&=-\frac{1}{N}\sum_{p=1}^\infty\Tr\left(U^{-p}\right)z^{p-1}\,,\qquad|z|<1\,.\label{eqn:R-_def}
\end{align}
The eigenvalue density $\rho(z)$ of a unitary matrix is given by
\begin{equation}
    \rho(z)=\sum_{n=1}^N\delta(z-\lambda_n),\qquad|z|=1\,,
\end{equation}
where $\lambda_n$ represents the $n$th eigenvalue of the unitary matrix $U$. This quantity can be related to the resolvents using the relation,
\begin{equation}
    \rho(z)=\frac{N}{2\pi i}\lim_{\xi\rightarrow z}\left(R^+(\xi)-R^-(1/\xi^\dagger)\right)\,,\qquad|z|=1\,.\label{eqn:rho_def_resolvents}
\end{equation}
Before taking the double-scaling limit, we first evaluate the correlators of these quantities in the finite-$N$ unitary matrix ensemble with generic parameters $t_l^\pm\in\mathbb{C}$, $t_l^+=(t_l^-)^*$.

\subsection{Correlators in the finite-$N$ unitary matrix ensemble}

The Fredholm determinant expansion in (\ref{eqn:Fredholm_det_expansion}) and (\ref{eqn:original_expression_GNm}) allows us to evaluate correlators of products of traces involving the unitary matrix $U$ by taking appropriate derivatives with respect to the parameters $t_p^\pm$ on both sides of the identity (\ref{eqn:Fredholm_det_expansion}).
We define the differential operators,
\begin{align}
    D^-(z)&\equiv-\frac{g}{N^2}\sum_{p=1}^\infty pz^{+p-1}\frac{\partial}{\partial t_p^-}\,,\qquad|z|<1\,,\\
    D^+(z)&\equiv+\frac{g}{N^2}\sum_{p=1}^\infty pz^{-p-1}\frac{\partial}{\partial t_p^+}\,,\qquad|z|>1\,,
\end{align}
such that
\begin{align}
    D^-(z)\,Z(N,g,t_l^\pm)&=\int_{U(N)}\dd U\,R^-(z)\exp\left[\frac{N}{g}\Tr\,W(U)\right]\,,\\
    \left(\frac{1}{z}+D^+(z)\right)\,Z(N,g,t_l^\pm)&=\int_{U(N)}\dd U\,R^+(z)\exp\left[\frac{N}{g}\Tr\,W(U)\right]\,.
\end{align}
More generally, we can obtain any $(n+1)$-point correlator involving $R^\pm(z)$ from the corresponding $n$-point correlator by acting with these operators as follows,
\begin{align}
    \braket{R^-(z)\prod_{i=1}^n\mathcal{O}_i}&=\frac{1}{Z(N,g,t_l^\pm)}D^-(z)\left(Z(N,g,t_l^\pm)\braket{\prod_{i=1}^n\mathcal{O}_i}\right)\,,\\
    \braket{R^+(z)\prod_{i=1}^n\mathcal{O}_i}&=\frac{1}{Z(N,g,t_l^\pm)}\left(\frac{1}{z}+D^+(z)\right)\left(Z(N,g,t_l^\pm)\braket{\prod_{i=1}^n\mathcal{O}_i}\right)\,.
\end{align}
Here, the $n$-point correlator is defined as,
\begin{equation}
    \braket{\prod_{i=1}^n\mathcal{O}_i}\equiv\frac{1}{Z(N,g,t_l^\pm)}\int_{U(N)}\dd U\left(\prod_{i=1}^n\mathcal{O}_i\right)\exp\left[\frac{N}{g}\Tr\,W(U)\right]\,.
\end{equation}
We also define the following quantities,
\begin{equation}
    R_0^-(z)\equiv-\frac{1}{g}\sum_{p=1}^\infty t_p^+z^{p-1}\,,\qquad R_0^+(z)=\frac{1}{z}+\frac{1}{g}\sum_{p=1}^\infty t_p^-z^{-p-1}\,,
\end{equation}
which represent the large-$N$ limits of the resolvents, as can be inferred from the identities,
\begin{align}
    D^-(z)Z^{(0)}(N,g,t_l^\pm)&=R_0^-(z)Z^{(0)}(N,g,t_l^\pm)\,,\\ \left(\frac{1}{z}+D^+(z)\right)Z^{(0)}(N,g,t_l^\pm)&=R_0^+(z)Z^{(0)}(N,g,t_l^\pm)\,.
\end{align}
For convenience, we introduce the definition,
\begin{multline}
    G_{N,\,n}^{(m)}(z_1,\dots,z_n)\equiv(-1)^{n+m}\frac{1}{m!}\prod_{i=1}^m\oint_{r_-}\frac{\dd u_i}{2\pi i}\,\frac{1}{m!}\prod_{j=1}^m\oint_{r_+}\frac{\dd v_j}{2\pi i}\,\frac{\prod_{1\leq i<j\leq m}(u_i-u_j)^2(v_i-v_j)^2}{\prod_{1\leq i,j\leq m}(u_i-v_j)^2}\prod_{i=1}^m\left(\frac{u_i}{v_i}\right)^N\\
    \times\prod_{i=1}^m\exp\left[\frac{N}{g}\sum_{l=1}^\infty\left(\frac{t_l^-}{l}\left(u_i^l-v_i^l\right)-\frac{t_l^+}{l}\left(u_i^{-l}-v_i^{-l}\right)\right)\right]\prod_{q=1}^n\left(\sum_{j=1}^m\left(\frac{1}{z_q-u_j^{-1}}-\frac{1}{z_q-v_j^{-1}}\right)\right)\,,\label{eqn:GNmCorrelators}
\end{multline}
which satisfies,\footnote{Here, we are identifying the $n=0$ terms as the terms $G_N^{(m)}$ of the Fredholm determinant expansion. Note that the argument $z_{n+1}$ must have absolute value larger than $1/r_-$ if it is associated with $D^+$ and must have absolute value smaller than $1/r_+$ if it is associated with $D^-$. The expression (\ref{eqn:GNmCorrelators}) admits any arguments $z_q$ provided they satisfy $|z_q|\not\in(1/r_+,\,1/r_-)$.}
\begin{equation}
    D^\pm(z_{n+1})G_{N,\,n}^{(m)}(z_1,\dots,z_n)=\pm \frac{1}{N}G_{N,\,n+1}^{(m)}(z_1,\dots z_{n+1})\,.
\end{equation}
Finally, using (\ref{eqn:rho_def_resolvents}), we see that the operator corresponding to an insertion of $\rho(z)$ is given by,
\begin{equation}
    D^\rho(z)=\frac{N}{2\pi i}\lim_{\xi\rightarrow z}\left(\frac{1}{\xi}+D^+(\xi)-D^-(1/\xi^*)\right)\,,\qquad |z|=1\,,
\end{equation}
where it is understood that the limit is taken after the application of the operator associated with $\xi$.

\subsubsection{The one-point function}

Before investigating higher-point functions, we will study the case of a single insertion of $\rho(z)$. We have shown that this is given by
\begin{equation}
    \braket{\rho(z)}=\frac{1}{Z(N,g,t_l^\pm)}D^\rho(z)Z(N,g,t_l^\pm)\,.
\end{equation}
There are two types of contributions to the one-point function. The large-$N$ contribution comes from the differential operator acting on $Z^{(0)}(N,g,t_l^\pm)$, while the finite-$N$ corrections come from the operator acting on the terms in the Fredholm determinant expansion. Concretely, we find,
\begin{align}
    \braket{\rho(z)}=\rho_1^{(0)}(z)+\sum_{m=1}^\infty\rho_1^{(m)}(z)\,,
\end{align}
where $\rho_1^{(0)}(z)$ represents the large-$N$ eigenvalue density,
\begin{equation}
    \rho_1^{(0)}(z)=\frac{N}{2\pi iz}\left(1+\frac{1}{g}\sum_{p=1}^\infty\left(t_p^+z^p+t_p^-z^{-p}\right)\right)\,,\qquad|z|=1\,,\label{eqn:525}
\end{equation}
while $\rho_1^{(m)}(z)$ represents the contribution coming from the $m$th term in the Fredholm determinant expansion,\footnote{Importantly, the limit should be taken after evaluating the integrals involved in the expression inside the parenthesis. These integrals should be evaluated for $|\xi|>\max(r_+,\,1/r_-)$.}
\begin{equation}
    \rho_1^{(m)}(z)=\frac{1}{2\pi i}\frac{Z^{(0)}(N,g,t_l^\pm)}{Z(N,g,t_l^\pm)}\lim_{\xi\rightarrow z}\left(G^{(m)}_{N,\,1}(\xi)+G^{(m)}_{N,\,1}(1/\xi^*)\right)\,,\qquad|z|=1\,.
\end{equation}
\subsubsection{The two-point function}
The two-point function can be obtained by applying the operator $D^\rho(z)$ to the one-point function of $\rho(y)$,
\begin{equation}
    \braket{\rho(z)\rho(y)}=\frac{1}{Z(N,g,t_l^\pm)}D^\rho(z)\left(Z(N,g,t_l^\pm)\braket{\rho(y)}\right)\,.
\end{equation}
For clarity, we separate the right side of this identity into five contributions which we evaluate independently. We identify the following contributions.
\begin{itemize}
\item From two copies of the large-$N$ eigenvalue density $\rho_1^{(0)}(z)$ and $\rho_1^{(0)}(y)$:
\begin{equation}
    \rho_1^{(0)}(y)\times\frac{1}{Z^{(0)}(N,g,t_l^\pm)}D^\rho(z)Z^{(0)}(N,g,t_l^\pm)=\rho_1^{(0)}(z)\rho_1^{(0)}(y)\,.
\end{equation}
\item From the finite-$N$ correction terms $\rho_1^{(m)}(z)$ and the large-$N$ eigenvalue density $\rho_1^{(0)}(y)$:
\begin{equation}
    \rho_1^{(0)}(y)\times\frac{Z^{(0)}(N,g,t_l^\pm)}{Z(N,g,t_l^\pm)}\frac{N}{2\pi i}\lim_{\xi\rightarrow z}\left(D^+(\xi)-D^-(1/\xi^*)\right)\sum_{m=1}^\infty G_N^{(m)}=\sum_{m=1}^\infty\rho_1^{(m)}(z)\rho_1^{(0)}(y)\,.
\end{equation}
\item From the connected part of the correlator associated with the large-$N$ limit:
\begin{equation}
    \frac{N}{2\pi i}\lim_{\xi\rightarrow z}\left(D^+(\xi)-D^-(1/\xi^*)\right)\rho^{(0)}(y)=\frac{1}{(2\pi i)^2}\times\frac{2}{(z-y)^2}\,.\label{eqn:530}
\end{equation}
\item From the large-$N$ eigenvalue density $\rho_1^{(0)}(z)$ and the finite-$N$ correction terms $\rho_1^{(m)}(y)$:
\begin{equation}
    \sum_{m=1}^\infty\rho_1^{(m)}(y)\times\frac{1}{Z^{(0)}(N,g,t_l^\pm)}D^\rho(z)Z^{(0)}(N,g,t_l^\pm)=\sum_{m=1}^\infty\rho_1^{(0)}(z)\rho_1^{(m)}(y)\,.
\end{equation}
\item From the connected part of the correlator associated with finite-$N$ corrections:
\begin{multline}
    \frac{Z^{(0)}(N,g,t_l^\pm)}{Z(N,g,t_l^\pm)}\frac{N}{2\pi i}\lim_{\xi\rightarrow z}\left(D^+(\xi)-D^-(1/\xi^*)\right)\sum_{m=1}^\infty\frac{Z(N,g,t_l^\pm)}{Z^{(0)}(N,g,t_l^\pm)}\rho_1^{(m)}(y)=\frac{1}{(2\pi i)^2}\frac{Z^{(0)}(N,g,t_l^\pm)}{Z(N,g,t_l^\pm)}\\
    \times\lim_{\xi\rightarrow z}\lim_{\xi'\rightarrow y}\sum_{m=1}^\infty\left(G_{N,\,2}^{(m)}(\xi',\xi)+G_{N,\,2}^{(m)}(1/\xi'{}^*,\xi)+G_{N,\,2}^{(m)}(\xi',1/\xi^*)+G_{N,\,2}^{(m)}(1/\xi'{}^*,1/\xi^*)\right)\,.\label{eqn:532}
\end{multline}
\end{itemize}
In the following, we will use the notation $\rho^{(0)}_2(z,y)$ to refer to the right side of (\ref{eqn:530}) and the notation $\rho^{(m)}_2
(z,y)$ to refer to the $m$th term in the summation on the right side of (\ref{eqn:532}), including the factors of $\frac{1}{(2\pi i)^2}$ and $\frac{Z^{(0)}(N,g,t_l^\pm)}{Z(N,g,t_l^\pm)}$. We will define the quantities $\rho^{(m)}_n(z_1,\dots,z_n)$ analogously, including the factors of $\frac{1}{(2\pi i)^n}$ and $\frac{Z^{(0)}(N,g,t_l^\pm)}{Z(N,g,t_l^\pm)}$.\footnote{There is no higher-point analog $\rho_n^{(0)}(z_1,\dots,z_n)$ for the connected part of the correlator associated with the large-$N$ limit.}\\\\
Adding all of the five individual contributions, we find the following expression for the two-point function:
\begin{equation}
    \braket{\rho(z)\rho(y)}=\left(\rho_1^{(0)}(z)\rho_1^{(0)}(y)+\rho^{(0)}_2(z,y)\right)+\sum_{m=1}^\infty\left(\rho_1^{(m)}(z)\rho^{(0)}_1(y)+\rho^{(0)}_1(z)\rho^{(m)}_1(y)+\rho^{(m)}_2(z,y)\right)\,.
\end{equation}
\subsubsection{The higher-point functions}
We obtain the $n$-point functions $\braket{\rho(z_1)\dots\rho(z_n)}$ inductively using the operator $D^\rho(z)$:
\begin{equation}
    \braket{\rho(z_1)\dots\rho(z_n)}=\frac{1}{Z(N,g,t_l^\pm)}D^\rho(z_n)\left(Z(N,g,t_l^\pm)\braket{\rho(z_1)\dots\rho(z_{n-1})}\right)\,.
\end{equation}
These will take the form of a sum over sectors labeled by the non-negative integer $m$. The $m=0$ sector consists of terms which are given by products of the large-$N$ one-point functions $\rho^{(0)}_1(z_i)$ and large-$N$ connected two-point correlators $\rho_2^{(0)}(z_j,z_k)$. The factors in each such term correspond to distinct coordinates $z_i$ among $z_1,\dots,z_n$, and each of the $n$ coordinates $z_1,\dots,z_n$ is assigned to one of the factors. We will use the notation $\braket{\rho(z_1)\dots\rho(z_n)}^{(0)}$ to refer to the $m=0$ sector of the $n$-point function. For the first few values of $n$, this takes the form,
\begin{align}
    \braket{\rho(z_1)}^{(0)}&=\rho_1^{(0)}(z_1)\,,\\
    \braket{\rho(z_1)\rho(z_2)}^{(0)}&=\rho_1^{(0)}(z_1)\rho_1^{(0)}(z_2)+\rho_2^{(0)}(z_1,z_2)\,,\\
    \braket{\rho(z_1)\rho(z_2)\rho(z_3)}^{(0)}&=\rho_1^{(0)}(z_1)\rho_1^{(0)}(z_2)\rho_1^{(0)}(z_3)+\rho_2^{(0)}(z_1,z_2)\rho_1^{(0)}(z_3)+\rho_2^{(0)}(z_1,z_3)\rho_1^{(0)}(z_2)\nonumber\\
    &\qquad\qquad\qquad\qquad\qquad\qquad\qquad\qquad\qquad\qquad\qquad+\rho_2^{(0)}(z_2,z_3)\rho_1^{(0)}(z_1)\,.
\end{align}
The sector corresponding to a positive integer $m$ consists of terms that are products of the form
\begin{equation}
    \frac{1}{l!}\rho_l^{(m)}(z_{\sigma(1)},\dots z_{\sigma(l)})\times\frac{1}{(n-l)!}\braket{\rho(z_{\sigma(l+1)})\dots\rho(z_{\sigma(n)})}^{(0)}\,,
\end{equation}
where $1\leq l\leq n$ and $\sigma\in S_n$ represents a permutation of the first $n$ positive integers.\footnote{For the case $l=n$, this is simply $\rho_n^{(m)}(z_1,\dots,z_n)$.} The $n$-point functions therefore take the form,\footnote{Note that both $\rho_n^{(m)}(z_1,\dots,z_n)$ and $\braket{\rho(z_1)\dots\rho(z_n)}^{(0)}$ are symmetric in their arguments, which means that the sum over permutations in equation (\ref{eqn:539}) can be replaced with a sum over subsets of size $l$ of the set $\{z_1,\dots,z_n\}$, with the symmetry factors $1/l!$ and $1/(n-l)!$ dropped.}
\begin{multline}
    \braket{\rho(z_1)\dots\rho(z_n)}=\braket{\rho(z_1)\dots\rho(z_n)}^{(0)}\\
    +\sum_{m=1}^\infty\sum_{l=1}^n\sum_{\sigma\in S_n}\frac{1}{l!}\rho_l^{(m)}(z_{\sigma(1)},\dots z_{\sigma(l)})\times\frac{1}{(n-l)!}\braket{\rho(z_{\sigma(l+1)})\dots\rho(z_{\sigma(n)})}^{(0)}\,.\label{eqn:539}
\end{multline}
We now prove the induction step. Applying $D^\rho(z_n)$ to the $m=0$ sector of the $(n-1)$-point function, we find,
\begin{multline}
    \frac{1}{Z(N,g,t_l^\pm)}D^\rho(z_n)\left(Z(N,g,t_l^\pm)\braket{\rho(z_1)\dots\rho(z_{n-1})}^{(0)}\right)=\braket{\rho(z_1)\dots\rho(z_n)}^{(0)}\\
    +\sum_{m=1}^\infty\rho_1^{(m)}(z_n)\braket{\rho(z_1)\dots\rho(z_{n-1})}^{(0)}\,.\label{eqn:540}
\end{multline}
We have
\begin{equation}
    \frac{1}{Z^{(0)}(N,g,t_l^\pm)}D^\rho(z_n)\left(Z^{(0)}(N,g,t_l^\pm)\braket{\rho(z_1)\dots\rho(z_{n-1})}^{(0)}\right)=\braket{\rho(z_1)\dots\rho(z_n)}^{(0)}\,,
\end{equation}
Also,
\begin{multline}
    \frac{Z^{(0)}(N,g,t_l^\pm)}{Z(N,g,t_l^\pm)}\frac{N}{2\pi i}\lim_{\xi\rightarrow z_n}\left(D^+(\xi)-D^-(1/\xi^*)\right)\left(1+\sum_{m=1}^\infty G^{(m)}_N\right)\braket{\rho(z_1)\dots\rho(z_{n-1})}^{(0)}\\
    =\sum_{m=1}^\infty\rho^{(m)}_1(z_n)\braket{\rho(z_1)\dots\rho(z_{n-1})}^{(0)}\,.
\end{multline}
Applying $D^\rho(z_n)$ to one of the terms in an $m\geq1$ sector of the $(n-1)$-point function, we find,
\begin{multline}
    \frac{1}{Z(N,g,t_l^\pm)}D^\rho(z_n)\left(Z(N,g,t_l^\pm)\rho_l^{(m)}(z_{\sigma(1)},\dots,z_{\sigma(l)})\braket{\rho(z_{\sigma(l+1)})\dots\rho(z_{\sigma(n-1)})}^{(0)}\right)\\
    =\rho_l^{(m)}(z_{\sigma(1)},\dots,z_{\sigma(l)})\braket{\rho(z_{\sigma(l+1)})\dots\rho(z_{\sigma(n-1)})\rho(z_n)}^{(0)}\\
    +\rho_{l+1}^{(m)}(z_{\sigma(1)},\dots,z_{\sigma(l)},z_n)\braket{\rho(z_{\sigma(l+1)})\dots\rho(z_{\sigma(n-1)})}^{(0)}\,.\label{eqn:543}
\end{multline}
Combining (\ref{eqn:540}) and (\ref{eqn:543}), the identity (\ref{eqn:539}) follows. 

\subsection{Double-scaled correlators and their string theory interpretation}
Having obtained the exact form (\ref{eqn:539}) for the $n$-point correlator $\braket{\rho(z_1)\dots\rho(z_n)}$ in the finite-$N$ unitary matrix model, we now proceed to the double-scaling limit and its physical interpretation in minimal superstring theory.\\\\
First, consider the large-$N$ contribution, $\braket{\rho(z_1)\dots\rho(z_n)}^{(0)}$, which takes the form of a sum of terms which are products of large-$N$ one-point functions, $\rho_1^{(0)}(z_i)$, and large-$N$ two-point functions, $\rho_2^{(0)}(z_j,z_k)$. The expression (\ref{eqn:525}) for the one-point function can be written as,
\begin{equation}
    \rho_1^{(0)}(z)=\frac{N}{2\pi ig}\,y^+(z)\,.
\end{equation}
Performing the double-scaling procedure outlined in section \ref{sec:2.1}, and noting that
\begin{equation}
    v^{\pm}{}'(x)=4\Tilde{y}^\pm(x)\,,
\end{equation}
and that the double-scaled eigenvalue density is related to the one at finite $N$ via\footnote{The negative sign in this equation appears because the finite-$N$ density was defined such that $\rho(\theta)=\rho(z)\frac{\dd z}{\dd\theta}$ is positive. Since $\theta$ is measured counterclockwise, $\dd x$ would be negative near $z=-1$.}
\begin{equation}
    \Tilde{\rho}(x)=-\rho(z)\frac{\dd z}{\dd x}\,,
\end{equation}
we find that the double-scaled one-point function takes the form,
\begin{equation}
    \Tilde{\rho}_1^{(0)}(x)=\frac{1}{2\pi i}\kappa v^+{}'(x)\,.
\end{equation}
Multiplying by $2\pi i$, we identify this expression with the $x$-derivative of a single disk diagram supported by an $\nb$-boundary,
\begin{equation}
    2\pi i\,\Tilde{\rho}_1^{(0)}(x)=\partial_x\TT{\nb}{(x,\,1)}\,.
\end{equation}
Similarly, we compute the double-scaling limit of the large-$N$ two-point function and find that it takes the form,
\begin{equation}
    \Tilde{\rho}_2^{(0)}(x,y)=\frac{1}{(2\pi i)^2}\frac{2}{(x-y)^2}\,.
\end{equation}
Multiplying by $(2\pi i)^2$, we identify this expression with the mixed partial derivative of a single annulus diagram connecting two $\nb$-boundaries,
\begin{equation}
    (2\pi i)^2\,\Tilde{\rho}_2^{(0)}(x,y)=\partial_x \partial_y A^{\nb}_{(x,1), (y,1)}\,.
\end{equation}
Thus, we interpret the contribution, $(2\pi i)^n\braket{\Tilde{\rho}(x_1)\dots\Tilde{\rho}(x_n)}^{(0)}$ as the sum over all possible disk and annulus diagrams supported on $n$ $\nb$-boundaries associated with moduli $x_1,\dots,x_n$, followed by a partial derivative with respect to each of the moduli.\footnote{Note that in each of the allowed diagrams, each $\nb$-boundary bounds single diagram.} For instance, the large-$N$ one-point function corresponds to a single disk diagram. The two-point function corresponds to the sum of an annulus diagram and a product of two disk diagrams. The three-point function corresponds to a sum consisting of three diagrams involving the product of an annulus and a disk, and a diagram consisting of the product of three disks, and so on.\\\\
The finite-$N$ contributions appearing due to the terms in the Fredholm determinant expansion are more interesting. The double-scaling procedure allows us to find precise expressions for the terms $G_{N,\,n}^{(m)}(z_1,\dots,z_n)$ in (\ref{eqn:GNmCorrelators}). Their double-scaling limit satisfies,
\begin{equation}
    G_{N,\,n}^{(m)}(z_1,\dots,z_n)=\left(-\frac{k}{i\epsilon}\right)^n\times\mathcal{G}^{(m)}(\kappa;x_1,\dots,x_n)\,,
\end{equation}
where
\begin{multline}
    \mathcal{G}^{(m)}(\kappa;x_1,\dots,x_n)=\frac{1}{m!}\prod_{i=1}^m\int_{\mathcal{C}_+}\frac{\dd \lambda_i}{2\pi}\,\frac{1}{m!}\prod_{j=1}^m\int_{\mathcal{C}_-}\frac{\dd \sigma_j}{2\pi}\,\frac{\prod_{1\leq i<j\leq m}(\lambda_i-\lambda_j)^2(\sigma_i-\sigma_j)^2}{\prod_{1\leq i,j\leq m}(\lambda_i-\sigma_j)^2} \\
    \times\prod_{q=1}^n\left[\sum_{j=1}^m\left(\frac{1}{x_q-\lambda_j}-\frac{1}{x_q-\sigma_j}\right)\right]\times\exp\left[\kappa\sum_{i=1}^m\left(v^+(\lambda_i)+v^-(\sigma_i)\right)\right]\,.
\end{multline}
Thus, the contribution to the $n$-point function coming from the $m$th term in the Fredholm determinant expansion takes the form,
\begin{multline}
    \Tilde{\rho}_n^{(m)}(x_1,\dots,x_n)=\left(\frac{1}{2\pi i}\right)^n\frac{1}{\mathcal{Z}(\kappa)}\lim_{\zeta_q\rightarrow x_q}\frac{1}{m!}\prod_{i=1}^m\int_{\mathcal{C}_+}\frac{\dd \lambda_i}{2\pi}\,\frac{1}{m!}\prod_{j=1}^m\int_{\mathcal{C}_-}\frac{\dd \sigma_j}{2\pi}\,\frac{\prod_{1\leq i<j\leq m}(\lambda_i-\lambda_j)^2(\sigma_i-\sigma_j)^2}{\prod_{1\leq i,j\leq m}(\lambda_i-\sigma_j)^2} \\
    \times\prod_{q=1}^n\left[\sum_{j=1}^m\left(\frac{1}{\zeta_q-\lambda_j}-\frac{1}{\zeta_q-\sigma_j}+\frac{1}{\zeta_q^*-\lambda_j}-\frac{1}{\zeta_q^*-\sigma_j}\right)\right]\times\exp\left[\kappa\sum_{i=1}^m\left(v^+(\lambda_i)+v^-(\sigma_i)\right)\right]\,.
\end{multline}
Momentarily ignoring the issue of regularization and taking $\zeta_q=x_q\in\mathbb{R}$ in the integrand, we recognize the factors in the product over $q$ as the $x_q$-derivatives of the product of the annulus diagrams which connect an $\nb$-boundary with modulus $x_q$ to the $\nb$-brane with modulus $\lambda_j$ and to the anti-$\nb$-brane with modulus $\sigma_j$,
\begin{equation}
    \prod_{q=1}^n\left[\sum_{j=1}^m\left(\frac{2}{x_q-\lambda_j}-\frac{2}{x_q-\sigma_j}\right)\right]=\partial_{x_1}\dots\partial_{x_n}\prod_{q=1}^n\left[\sum_{j=1}^m\left(A^{\nb}_{(x_q,1), (\lambda_j,1)}+A^{\nb}_{(x_q,1), (\sigma_j,-1)}\right)\right]\,.
\end{equation}
Thus, we interpret the contribution to the $n$-point function coming from the $m$th term in the Fredholm determinant expansion, $(2\pi i)^n\mathcal{Z}(\kappa)\Tilde{\rho}_n^{(m)}(x_1,\dots,x_n)$, as the contribution of all diagrams involving annuli connecting the $\nb$-boundaries with moduli $x_1,\dots,x_n$ to the $\nb$-branes and anti-$\nb$-branes in all possible ways, followed by taking a partial derivative with respect to each of the moduli $x_1,\dots,x_n$.\\\\
Combining the contributions as we described in equation (\ref{eqn:539}), we find that, just like the partition function $\mathcal{Z}(\kappa)$, the quantity, $(2\pi i)^n\mathcal{Z}(\kappa)\braket{\Tilde{\rho}(x_1)\dots\Tilde{\rho}(x_n)}$, corresponds in the string theory language to a sum over sectors labeled by the nonnegative integer $m$. The contribution from the $m$th sector consists of all possible disk and annulus diagrams supported by the $\nb$-boundaries with moduli $x_1,\dots,x_n$. Each $\nb$-boundary can either support a disk diagram or an annulus diagram connecting it to another one of the $\nb$-boundaries or to one of the $m$ $\nb$-branes or to one of the $m$ anti-$\nb$-branes. The contribution from the $m$th sector also includes all disk and annulus diagrams supported on the $m$ $\nb$-branes and $m$ anti-$\nb$-branes, and all of these contributions are integrated over the moduli space of the fundamental branes. Thus, we conclude that inserting a copy of $2\pi i\Tilde{\rho}(x)$ into a correlator corresponds to inserting an $\nb$-boundary with modulus $x$ on the string theory side of the duality, followed by taking a derivative with respect to the modulus $x$ of the resulting sum over diagrams. An important point to note is that it is not the correlator $\braket{\Tilde{\rho}(x_1)\dots\Tilde{\rho}(x_n)}$ that admits a natural string theory interpretation, but rather, the correlator multiplied by the partition function $\mathcal{Z}(\kappa)$. It is this quantity that can be written as a sum over the distinct sectors of the string theory vacuum containing $m$ $\nb$-branes and $m$ anti-$\nb$-branes.
\section{Discussion and future directions}
\label{sec:conc}
In this work, we obtained the complete non-perturbative partition function of the $\mathcal{N}=1$ $(2,4k)$ minimal superstring theory with type 0B GSO projection in the ungapped phase, both indirectly by studying its dual matrix integral description, and directly by identifying the fundamental objects of the string theory. We provided a Hilbert space interpretation of the theory in terms of a quantum mechanical system of dressed free fermions which allowed us to interpret the partition function of the string theory as the Witten index of the fermionic system. We also found that all of these results generalize to the $k\rightarrow\infty$ limit of this family of theories, which corresponds to JT supergravity.\\\\
We close with a list of open questions that we believe merit pursuing. First, within the context of the $(2,4k)$ minimal superstring, we found evidence that the (anti-)$\nb$-branes are the fundamental objects of the theory and that the only non-vanishing diagrams supported by these should be the disk and annulus diagrams. It would be very interesting to understand why this is the case. Additionally, although our work provides the expression for the partition function of JT supergravity, it does so indirectly by appealing to the duality between this theory and the double-scaled matrix integral, or through its realization as the $k\rightarrow\infty$ limit of the $(2,4k)$ minimal superstring theory. It would be very interesting to derive this expression directly, perhaps using a supersymmetric localization argument, and to provide a concrete interpretation of the $\nb$-branes and their charge-conjugate partners in the language of dilaton gravity. It would also be worthwhile to generalize the ideas in the present work to other string theories that admit matrix model duals. Examples include the gapped phase of the type 0B minimal superstring theories, type 0A minimal superstring theories, the bosonic $(2,2k+1)$ minimal string theories and their large-$k$ (JT gravity) limit, the Virasoro minimal string \cite{Collier:2023cyw}, the complex Liouville string \cite{Collier:2024kmo, Collier:2024kwt, Collier:2024lys, Collier:2024mlg}, and the $c=1$ string theory. In contrast to the ungapped phase of type 0B minimal superstring theory for which the free energy is trivial at the perturbative level, in these theories, the free energy has a non-trivial genus expansion. This is a key difference that makes the analysis in these theories more complicated. \\\\
We also hope that our work will provide insight into non-perturbative contributions normally associated with D-brane instantons in theories that do not admit a matrix model description. Normally, contributions associated with instanton sectors are asymptotic series which have to be Borel-\'Ecalle-resummed in order to arrive at exact non-perturbative expressions. However, it would be very interesting to study whether these exact expressions could instead be obtained as contributions coming from a different set of branes, as we have found to be the case for the minimal superstring theory, where the instanton branes simply act as saddle points for the integral over the moduli space of the fundamental branes.

\vspace{2\baselineskip}
\paragraph{Acknowledgments.}

We thank Alexander Frenkel, Clifford Johnson, Raghu Mahajan, Marcos Mari\~no, Ricardo Schiappa, Maximilian Schwick, Stephen Shenker, and Xiaoliang Qi for helpful conversations and correspondence. C.M.~is supported by the DOE through DESC0013528 and the QuantISED grant DESC0020360.

\newpage

\appendix
\section{Unitary matrix integrals as Fredholm determinants}\label{app:FredProof}

In this appendix, we provide a brief sketch of the derivation of the Fredholm determinant expansion.
Consider the unitary matrix integral with single-trace potential,
\begin{equation}
    Z(N,g,t_l^\pm)=\int_{U(N)}\dd U\,\exp\left[\frac{N}{g}\Tr\,\sum_{l=1}^k\left(\frac{t_l^+}{l}U^l+\frac{t_l^-}{l}U^{-l}\right)\right]\,,\label{eqn:unitary_matrix_integral_appendix}
\end{equation}
where $t_l^\pm$ for $1\leq l\leq k$ are fixed complex parameters with $(t_l^+)^*=t_l^-$, $g \in \mathbb{R}^+$ is the 't Hooft coupling, and $N$ is a positive integer denoting the degree of the unitary group. The Fredholm determinant expansion takes the form
\begin{equation}
    \frac{Z(N,g,t_l^\pm)}{Z^{(0)}(N,g,t_l^\pm)}=1+\sum_{m=1}^\infty G_N^{(m)}\,,\label{eqn:Fredholm_det_expansion_appendix}
\end{equation}
where
\begin{multline}
    G_N^{(m)}=(-1)^m\frac{1}{m!}\prod_{i=1}^m\oint_{r_-}\frac{\dd u_i}{2\pi i}\,\frac{1}{m!}\prod_{j=1}^m\oint_{r_+}\frac{\dd v_j}{2\pi i}\,\frac{\prod_{1\leq i<j\leq m}(u_i-u_j)^2(v_i-v_j)^2}{\prod_{1\leq i,j\leq m}(u_i-v_j)^2}\,\prod_{i=1}^m\left(\frac{u_i}{v_i}\right)^N\\
    \times\prod_{i=1}^m\exp\left[\frac{N}{g}\sum_{l=1}^k\left(\frac{t_l^-}{l}\left(u_i^l-v_i^l\right)-\frac{t_l^+}{l}\left(u_i^{-l}-v_i^{-l}\right)\right)\right]\,. \label{eqn:final_form_GNm}
\end{multline}
Here, the contours for the eigenvalues $u_i$ are circles of radius $r_-\rightarrow 1^-$ and the contours for the anti-eigenvalues $v_j$ are circles of radius $r_+\rightarrow1^+$.\\\\
We now sketch the proof of the identity (\ref{eqn:Fredholm_det_expansion_appendix}), following Ref.~\cite{Eniceicu:2023uvd} which relies crucially on \cite{borodin2000fredholm, Geronimo:1979iy, Murthy:2022ien, tracy2013diagonal, tracy2017natural, Liu:2022olj}. The expression of a unitary matrix integral as a Fredholm determinant is a result of a theorem of Borodin and Okounkov \cite{borodin2000fredholm} which had been discovered independently by Geronimo and Case \cite{Geronimo:1979iy}.\\\\
The first step in deriving the expansion (\ref{eqn:Fredholm_det_expansion_appendix}) is expressing the unitary matrix integral (\ref{eqn:unitary_matrix_integral_appendix}) as the determinant of a Toeplitz matrix. A Toeplitz matrix is a matrix in which each diagonal is constant. Let us define
\begin{equation}
    \varphi(\xi)=\exp\left[\frac{N}{g}\sum_{l=1}^k\left(\frac{t_l^+}{l}\xi^l+\frac{t_l^-}{l}\xi^{-l}\right)\right]\,,
\end{equation}
in the neighborhood of the unit circle $|\xi|=1$.
The $j$th Fourier coefficient of $\varphi(\xi)$ is
\begin{equation}
    \varphi_j=\frac{1}{2\pi i}\oint\dd\xi\,\varphi(\xi)\xi^{-(j+1)}\,.
\end{equation}
The matrix $M$ whose elements are given by $M_{ij}=\varphi_{i-j}$ is Toeplitz, and its determinant equals the unitary matrix integral in (\ref{eqn:unitary_matrix_integral_appendix}),\footnote{See Ref.~\cite{Eniceicu:2023uvd} for details.}
\begin{equation}
    \det(M)=Z(N,g,t_l^\pm)\,.
\end{equation}
The Borodin-Okounkov-Geronimo-Case theorem then allows us to write the Toeplitz determinant as a Fredholm determinant, and one finds the following series expansion,
\begin{equation}
    \frac{Z(N,g,t_l^\pm)}{Z^{(0)}(N,g,t_l^\pm)}=1+\sum_{m=1}^\infty\frac{(-1)^m}{m!}\sum_{0\leq n_1,n_2,\dots,n_m}\det[f_N(n_i,n_j)]_{i,j=1,\dots,m}\,,\label{eqn:Fred_app}
\end{equation}
with
\begin{multline}
    f_N(n_i,n_j)=\oint_{r_-}\frac{\dd u}{2\pi i}\oint_{r_+}\frac{\dd v}{2\pi i}u^{n_i}v^{-n_j}\left(\frac{u}{v}\right)^{N+1}\frac{1}{1-u/v}\\
    \times\exp\left[\frac{N}{g}\sum_{l=1}^k\left(\frac{t_l^-}{l}\left(u^l-v^l\right)-\frac{t_l^+}{l}\left(u^{-l}-v^{-l}\right)\right)\right]\,.
\end{multline}
Here,
\begin{equation}
    Z^{(0)}(N,g,t_l^\pm)=\exp\left(\frac{N^2}{g^2}\sum_{l=1}^k\frac{t_l^+t_l^-}{l}\right)\,,
\end{equation}
denotes the large-$N$ limit of the unitary matrix integral, and the contours of integration for $u$ and $v$ are circles centered at the origin of radii $r_-<1$ and $r_+>1$.\\\\
The right side of (\ref{eqn:Fred_app}) is a Fredholm determinant associated with a particular operator $T_N$ on $\ell^2(\mathbb{N})$, the space of all square-summable complex sequences. The operator in question can be defined through its action on the orthonormal basis $\{\ket{n}\}_{n\in\mathbb{N}}$:
\begin{equation}
    T_N\ket{n}=\sum_{n'=0}^\infty f_N(n,n')\ket{n'}\,.
\end{equation}
The Fredholm determinant associated with $T_N$ with complex parameter $\mu$ is
\begin{equation}
    \det(\mathbb{1}-\mu T)=1+\sum_{m=1}^\infty(-\mu)^m\,\Tr\,\Lambda^m(T)\,,
\end{equation}
where
\begin{equation}
    \Tr\,\Lambda^m(T)=\frac{1}{m!}\sum_{0\leq n_1,\dots,n_m}\det[f_N(n_i,n_j)]_{i,j=1,\dots,m}\,.
\end{equation}
Thus, the right side of (\ref{eqn:Fred_app}) is $\det(\mathbb{1}-T)$.\\\\
Performing the sums over the integers $n_i$ in (\ref{eqn:Fred_app}), we obtain
\begin{align}
    \sum_{0\leq n_1,\dots,n_m}&\det\left[f_N(n_i,n_j)\right]_{i,j=1,\dots,m}=\prod_{i=1}^m\oint\frac{\dd u_i}{2\pi i u_i}\prod_{j=1}^m\oint\frac{\dd v_j}{2\pi i v_j}\prod_{i=1}^m\left(\frac{u_i}{v_i}\right)^{N+1}\prod_{i=1}^m\frac{1}{1-u_i/v_i}\times\nonumber\\
    &\times\det\left(\frac{1}{1-u_i/v_j}\right)_{i,j=1,\dots,m}\exp\left[\frac{N}{g}\sum_{l=1}^k\left(\frac{t_l^-}{l}\sum_{i=1}^m\left(u_i^l-v_i^l\right)-\frac{t_l^+}{l}\sum_{i=1}^m\left(u_i^{-l}-v_i^{-l}\right)\right)\right]\,.
\end{align}
The final observation which is provided by Refs.~\cite{tracy2013diagonal,tracy2017natural} is that one can perform the replacement
\begin{equation}
    \prod_{i=1}^m\frac{1}{1-u_i/v_i}\det\left[\frac{1}{1-u_i/v_j}\right]_{i,j=1,\dots,m}\rightarrow\frac{1}{m!}\det\left[\frac{1}{1-u_i/v_j}\right]^2_{i,j=1,\dots,m}\,,
\end{equation}
by symmetrizing over the variables $u_i$; see Ref.~\cite{Eniceicu:2023uvd} for details. Using the Cauchy determinant identity,
\begin{equation}
    \det\left[\frac{1}{v_j-u_i}\right]_{i,j=1,\dots,m}=\frac{\prod_{1\leq j<i\leq m}(u_j-u_i)(v_i-v_j)}{\prod_{i,j=1}^m(v_j-u_i)}\,,
\end{equation}
one arrives at the Fredholm determinant expansion (\ref{eqn:Fredholm_det_expansion_appendix}), with $G_N^{(m)}$ given by (\ref{eqn:final_form_GNm}).
\section{Pure supergravity}\label{app:pure_sugra}

In this appendix, we review our results in the particular case of pure supergravity which corresponds to the $(2,4)$ minimal superstring i.e., $k=1$. In this case, the potential takes the form,
\begin{equation}
    v^\pm(z)=\pm\frac{2i}{3}(z^3+6z)\,.
\end{equation}
It is expected that \cite{Marino:2008ya,Ahmed:2017lhl} once one makes the identification $c\equiv2^{5/3}\kappa^{2/3}$, the quantity,
\begin{equation}
    \mu(c)\equiv\sqrt{-F''(c)}\,,
\end{equation}
should satisfy the Painlev\'e II differential equation,
\begin{equation}
    \mu''(c)=2\mu(c)^3+c\mu(c)\,.
\end{equation}
Among the solutions of this equation, there exists a distinguished one called the Hastings-McLeod solution which is real for $c\in\mathbb{R}$, has no poles on the real axis, and has the asymptotic behavior $\mu(c)\sim\Ai(c)$, as $c\rightarrow\infty$ \cite{rosales1978similarity,hastings1980boundary}. In the following, we will show that the expressions we have found for the partition function indeed imply that $\mu(c)$ is the Hastings-McLeod solution of the Painlev\'e II equation.\\\\
Recall from section \ref{sec:3.3} that the momentum space expressions for the connected diagrams are
\begin{equation}
    D_{n|n}(\kappa)=-\frac{1}{n}\prod_{j=1}^{2n}\int_0^\infty\dd p_j\,\exp\left(-\epsilon\sum_{j=1}^{2n}p_j\right)\prod_{j=1}^{2n}g_-(p_j+p_{j+1})\,.
\end{equation}
We can explicitly evaluate the functions $g_-(P)$ representing the Fourier transform of the exponential of the potential,
\begin{equation}
    g_-(P)=e^{-2\kappa\epsilon}\int_{-\infty}^{+\infty}\frac{\dd x}{2\pi}\,\exp\left[-\frac{2i\kappa}{3}\left(x-\frac{i\epsilon}{2}\right)^3-i\left(P+4\kappa\right) x\right]\,.
\end{equation}
We restrict to the limit $\epsilon\to 0$, so that the Fourier transform above becomes the Airy function,
\begin{equation}
    g_-(P)=\int_{-\infty}^{+\infty}\frac{\dd x}{2\pi}\,\exp\left[-\frac{2i\kappa}{3}x^3-i\left(P+4\kappa\right) x\right]=\frac{1}{(2\kappa)^{1/3}}\Ai\left(c+(2\kappa)^{-1/3}P\right)\,.
\end{equation}
Thus, the connected diagrams are given by
\begin{equation}
    D_{n|n}(c)=-\frac{1}{n}\prod_{j=1}^{2n}\int_0^\infty\dd q_j\,\prod_{j=1}^{2n}\Ai\left(c+q_j+q_{j+1}\right)\,,
\end{equation}
with the identification $u_{2n+1}\equiv u_1$. In particular, the expression for the first connected diagram is
\begin{equation}
    D_{1|1}(c)=-\int_0^\infty\dd q\,q\,\Ai(c+q)^2\,,
\end{equation}
which can be evaluated in closed form using the defining property of the Airy function, $\text{Ai}''(x)=x\,\text{Ai}(x)$.
We find,
\begin{equation}
    D_{1|1}(c)=-\frac{1}{3}\left(2c^2\text{Ai}(c)^2-\text{Ai}(c)\text{Ai}'(c)-2c\text{Ai}'(c)^2\right)\,.
\end{equation}
For $\kappa=0$, this evaluates to $D_{1|1}(0)=- \frac{1}{6\pi\sqrt{3}}$ and as $\kappa\rightarrow\infty$, this approaches $0$.\\\\
In order to study the quantity $\mu(c)$, we first introduce several helpful definitions and identities. For $n\geq2$, we define
\begin{align}
    L_n(c)&\equiv\prod_{j=1}^{n-1}\int_0^\infty\dd q_j\,\Ai(c+q_1)\prod_{j=2}^n\Ai(c+q_{j-1}+q_j)\,,\\
    M_n(c)&\equiv\prod_{j=1}^{n-1}\int_0^\infty\dd q_j\,\Ai'(c+q_1)\prod_{j=2}^n\Ai(c+q_{j-1}+q_j)\,,\\
    N_n(c)&\equiv\prod_{j=1}^{n-1}\int_0^\infty\dd q_j\,\Ai''(c+q_1)\prod_{j=2}^n\Ai(c+q_{j-1}+q_j)\,.
\end{align}
It is worth mentioning that we have set $q_n=0$ to obtain these expressions. For $n=1$, we define
\begin{equation}
    L_1(c)=\Ai(c)\,,\qquad M_1(c)=\Ai'(c)\,,\qquad N_1(c)=\Ai''(c)\,.
\end{equation}
We can evaluate the first derivative of $L_n(c)$ by pairing the terms where the derivative acts on adjacent factors and using integration by parts to obtain
\begin{align}
    L_{2n}'(c)&=-\sum_{\substack{i,\,j\text{ odd}\\i+j=2n}}L_i(c)\,L_j(c)\,,\label{eqn:B14}\\
    L_{2n-1}'(c)&=M_{2n-1}(c)-\sum_{\substack{i\text{ even, }j\text{ odd}\\i+j=2n-1}}L_i(c)\,L_j(c)\,.
\end{align}
The first derivative of $M_n(c)$ can be obtained similarly,
\begin{align}
    M_{2n}'(c)&=-\sum_{\substack{i,\,j\text{ odd}\\i+j=2n}}M_i(c)\,L_j(c)\,,\\
    M_{2n-1}'(c)&=N_{2n-1}(c)-\sum_{\substack{i\text{ even, }j\text{ odd}\\i+j=2n-1}}\,M_i(c)\,L_j(c)\,.
\end{align}
We can also use integration by parts to express $M_{2n}(c)$ in terms of $L_j(c)$,
\begin{equation}
    M_{2n}(c)=\frac{1}{2}\sum_{\substack{i,\,j\text{ even}\\i+j=2n}}L_i(c)\,L_j(c)-\frac{1}{2}\sum_{\substack{i,\,j\text{ odd}\\i+j=2n}}L_i(c)\,L_j(c)\,.
\end{equation}
Finally, we express $N_{2n-1}(c)$ in terms of $L_j(c)$. The standard expression for the Airy kernel is
\begin{equation}
    \int_0^\infty\dd q\,\Ai(x+q)\Ai(y+q)=\frac{\Ai(x)\Ai'(y)-\Ai(y)\Ai'(x)}{x-y}\,,\qquad\text{ for }x\neq y\,.\label{eqn:B19}
\end{equation}
Since $\Ai''(x)=x\,\Ai(x)$, we have
\begin{equation}
    N_{2n-1}(c)=c\,L_{2n-1}(c)+\prod_{j=1}^{2n-2}\int_0^\infty\dd q_j\,q_1\,\Ai(c+q_1)\times\prod_{j=2}^{2n-2}\Ai(c+q_{j-1}+q_j)\times\Ai(c+q_{2n-2})\,.\label{eqn:B20}
\end{equation}
We can write
\begin{equation}
    q_1=(q_1-q_3)+(q_3-q_5)+\dots+(q_{2n-5}-q_{2n-3})+q_{2n-3}\,,
\end{equation}
in order to express (\ref{eqn:B20}) as a sum of terms which we can individually evaluate using (\ref{eqn:B19}). For example, by first evaluating the integral over $q_2$, the first term becomes
\begin{multline}
    \prod_{j=1}^{2n-2}\int_0^\infty\dd q_j\,(q_1-q_3)\,\Ai(c+q_1)\times\prod_{j=2}^{2n-2}\Ai(c+q_{j-1}+q_j)\times\Ai(c+q_{2n-2})\\
    =L_2(c)\,M_{2n-3}(c)-M_2(c)\,L_{2n-3}(c)\,.
\end{multline}
The second term can be obtained by first evaluating the integral over $q_4$, and so on. In the end, we find
\begin{equation}
    N_{2n-1}(c)=c\,L_{2n-1}(c)+\sum_{\substack{i\text{ even, }j\text{ odd}\\i+j=2n-1}}(L_i(c)\,M_j(c)-M_i(c)\,L_j(c))\,.
\end{equation}
Finally, using the previous identities, we find that the second derivative of $L_{2n-1}(c)$ takes the form,
\begin{equation}
    L_{2n-1}''(c)=M_{2n-1}'(c)-\sum_{\substack{i\text{ even, }j\text{ odd}\\i+j=2n-1}}L_i'(c)\,L_j(c)-\sum_{\substack{i\text{ even, }j\text{ odd}\\i+j=2n-1}}L_i(c)\,L_j'(c)\,.
\end{equation}
The first term becomes
\begin{equation}
\begin{split}
    M_{2n-1}'(c)&=N_{2n-1}(c)-\sum_{\substack{i\text{ even, }j\text{ odd}\\i+j=2n-1}}\,M_i(c)\,L_j(c)\\
    &=c\,L_{2n-1}(c)+\sum_{\substack{i\text{ even, }j\text{ odd}\\i+j=2n-1}}(L_i(c)\,M_j(c)-2M_i(c)\,L_j(c))\\
    &=c\,L_{2n-1}(c)+\sum_{\substack{i\text{ even, }j\text{ odd}\\i+j=2n-1}}L_i(c)\,M_j(c)+\sum_{\substack{j,\,k,\,l\text{ odd}\\j+k+l=2n-1}}L_j(c)\,L_k(c)\,L_l(c)\\
    &\qquad\qquad\qquad\qquad\qquad\qquad\qquad\qquad\qquad-\sum_{\substack{j\text{ odd, }k,\,l\text{ even}\\j+k+l=2n-1}}L_j(c)\,L_k(c)\,L_l(c)\,,
\end{split}
\end{equation}
the second term becomes
\begin{equation}
    -\sum_{\substack{i\text{ even, }j\text{ odd}\\i+j=2n-1}}L_i'(c)\,L_j(c)=\sum_{\substack{j,\,k,\,l\text{ odd}\\j+k+l=2n-1}}L_j(c)\,L_k(c)\,L_l(c)\,,
\end{equation}
and the third term becomes
\begin{equation}
    -\sum_{\substack{i\text{ even, }j\text{ odd}\\i+j=2n-1}}L_i(c)\,L_j'(c)=-\sum_{\substack{i\text{ even, }j\text{ odd}\\i+j=2n-1}}L_i(c)\,M_j(c)+\sum_{\substack{i,\,j\text{ even, }k\text{ odd}\\i+j+k=2n-1}}L_i(c)\,L_j(c)\,L_k(c)\,.\nonumber
\end{equation}
Summing these terms, we obtain
\begin{equation}
    L_{2n-1}''(c)=c\,L_{2n-1}(c)+2\sum_{\substack{j,\,k,\,l\text{ odd}\\j+k+l=2n-1}}L_j(c)\,L_k(c)\,L_l(c)\,.\label{eqn:B25}
\end{equation}
We can now evaluate $\mu(c)$ and its second derivative. The first derivative of the connected diagram $D_{n|n}(c)$ can be evaluated using integration by parts,
\begin{equation}
    D_{n|n}'(c)=L_{2n}(c)\,.
\end{equation}
The second derivative of the connected diagram $D_{n|n}(c)$ can then be evaluated using (\ref{eqn:B14}),
\begin{equation}
    D_{n|n}''(c)=-\sum_{\substack{i,\,j\text{ odd}\\i+j=2n}}L_i(c)\,L_j(c)\,
\end{equation}
The second derivative of the free energy can therefore be expressed as,
\begin{equation}
    F''(c)=\sum_{n=1}^\infty D_{n|n}''(c)=-\sum_{n=1}^\infty\sum_{\substack{i,\,j\text{ odd}\\i+j=2n}}L_i(c)\,L_j(c)=-\left(\sum_{i\text{ odd}}L_i(c)\right)^2\,,
\end{equation}
and we find,
\begin{equation}
    \mu(c)=\sum_{n=1}^\infty L_{2n-1}(c)\,.\label{eqn:B29}
\end{equation}
Using (\ref{eqn:B25}), the second derivative of $\mu(c)$ evaluates to
\begin{align}
    \mu''(c)&=\sum_{n=1}^\infty\left(c\,L_{2n-1}(c)+2\sum_{\substack{j,\,k,\,l\text{ odd}\\j+k+l=2n-1}}L_j(c)\,L_k(c)\,L_l(c)\right)\nonumber\\
    &=c\sum_{n=1}^\infty L_{2n-1}(c)+2\left(\sum_{j\text{ odd}}L_j(c)\right)^3\nonumber\\
    &=c\,\mu(c)+2\mu(c)^3\,,
\end{align}
which is exactly the form of the Painlev\'e II equation. Moreover, the first term in expression (\ref{eqn:B29}) is $L_1(c)=\Ai(c)$. Since the remaining terms are exponentially suppressed as $c\rightarrow\infty$ due to the nature of the asymptotic behavior of the Airy function, it follows that $\mu(c)\sim\Ai(c)$ as $c\rightarrow\infty$. Thus, we conclude that $\mu(c)$ is indeed the Hastings-McLeod solution of the Painlev\'e II equation.\\\\
We can also verify that the partition function is the Fredholm determinant,
\begin{equation}
    \mathcal{Z}(\kappa)=\det(\mathbb{1}-T)\,,
\end{equation}
where $T$ is the integral operator associated with the Airy kernel:
\begin{align}
    f(p,p')&=\braket{p'|T|p}=\theta(p)\theta(p')\int_{\mathcal{C}_+}\frac{\dd\lambda}{2\pi}\int_{\mathcal{C}_-}\frac{\dd\sigma}{2\pi}\,e^{ip\lambda}e^{-ip'\sigma}\frac{i}{\lambda-\sigma}\exp\left[\kappa(v^+(\lambda)+v^-(\sigma))\right]\nonumber\\
    &=\frac{1}{(2\kappa)^{1/3}}\theta(p)\theta(p')\int_0^\infty\dd q\,\Ai\left(c+(2\kappa)^{-1/3}p+q\right)\,\Ai\left(c+(2\kappa)^{-1/3}p'+q\right)\,.
\end{align}
If we modify the momentum space representation by rescaling $\Tilde{p}=(2\kappa)^{-1/3}p$, then
\begin{equation}
    T\ket{\Tilde{p}}=\int_{-\infty}^{+\infty}\dd\Tilde{p}'\,\Tilde{f}(\Tilde{p},\Tilde{p}')\ket{\Tilde{p}'}\,,
\end{equation}
with
\begin{equation}
    \Tilde{f}(\Tilde{p},\Tilde{p}')=\frac{\dd p'}{\dd\Tilde{p}'}\,f(p,p')\,.
\end{equation}
The kernel therefore takes the form
\begin{equation}
    \Tilde{f}(\Tilde{p},\Tilde{p}')=\braket{\Tilde{p}'|T|\Tilde{p}}=\theta(\Tilde{p})\theta(\Tilde{p}')\frac{\Ai\left(c+\Tilde{p}\right)\Ai'\left(c+\Tilde{p}'\right)-\Ai\left(c+\Tilde{p}'\right)\Ai'\left(c+\Tilde{p}\right)}{\Tilde{p}-\Tilde{p}'}\,.
\end{equation}
\section{JT supergravity}
\label{app:SJT}

In this appendix, we will focus on the case where $k\rightarrow\infty$ which corresponds to $\mathcal{N}=1$ JT supergravity \cite{Stanford:2019vob,Johnson:2021owr}. In the first part, we explain how the spectral curve of this theory is recovered by taking the $k\rightarrow\infty$ limit of the spectral curve associated with the $(2,4k)$ minimal superstring theory. We then prove that the terms $\mathcal{G}^{(m)}(\kappa)$ in the expression for the complete non-perturbative partition function of the theory are convergent, independent of the choices of contours, and real. The argument in section \ref{sec:4} continues to hold, and hence, we conclude that the expression we find for the partition function is convergent and expressible as the Fredholm determinant associated with a trace-class operator on $L^2(\mathbb{R})$.

\subsection{JT supergravity as the $k\rightarrow\infty$ limit of the minimal superstring theory}

Recall that the spectral curve associated with the $\mathcal{N}=1$ $(2,4k)$ minimal superstring theory with type 0B GSO projection in the ungapped phase takes the form,
\begin{equation}
    \Tilde{y}^\pm(k;x)=\pm i(-1)^k\,T_{2k}\left(\frac{ix}{2k}\right)\,,
\end{equation}
where $T_n(z)$ denotes the Chebyshev polynomial of the first kind defined by
\begin{equation}
\label{eqn:chebyshev}
    T_n(\cos(\theta))=\cos(n\theta)\,.
\end{equation}
The eigenvalue density associated with the spectral curve is supported on the real line $x\in\mathbb{R}$, and takes the form,
\begin{equation}
    \Tilde{\rho}(x)\;\propto\;i\left(\Tilde{y}^-(x)-\Tilde{y}^+(x)\right)\,.
\end{equation}
In the ungapped phase, both in the case of the minimal superstring theories and in the case of JT supergravity, the eigenvalue density covers the entire real line.\footnote{Conversely, the gapped phase of these theories has an associated eigenvalue density supported on the union of half-lines $(-\infty,-a]\cup[a,\infty)$, for some $a>0$.} For JT supergravity one expects
\begin{equation}
    \Tilde{\rho}_{\text{super-JT}}(x)\;\propto\;\cosh(x)\,.\label{eqn:rho_SJT_prediction}
\end{equation}
Indeed, upon setting
\begin{equation}
    \theta=-\frac{\pi}{2}+\frac{ix}{2k}\,,
\end{equation}
in (\ref{eqn:chebyshev}) with $n=2k$, we find
\begin{equation}
    (-1)^k\;\,T_{2k}\left(\sin \left(\frac{ix}{2k}\right)\right)=\cosh(x)\,.
\end{equation}
Taking the $k\to\infty$ limit, the spectral curve of the $(2,4k)$ minimal superstring theory becomes
\begin{equation}
    \Tilde{y}_{\text{super-JT}}^\pm(x)=\lim_{k\rightarrow\infty}\Tilde{y}^\pm(k;x)=\pm i\cosh(x)\,,
\end{equation}
from which we recover (\ref{eqn:rho_SJT_prediction}). Therefore, the effective potentials for JT supergravity in our conventions take the form,
\begin{equation}
    v^\pm(x)=4\int_0^x\dd\xi\;\Tilde{y}^\pm_{\text{super-JT}}(\xi)=\pm4i\sinh(x)\,.
\end{equation}

\subsection{Properties of the JT supergravity partition function}

In this subsection, we will show that the expression, 
\begin{equation}
    \mathcal{Z}(\kappa)=1+\sum_{m=1}^\infty\mathcal{G}^{(m)}(\kappa)\,,\label{eqn:ZSJTexpression}
\end{equation}
we obtained for the complete non-perturbative partition function of JT supergravity is convergent, independent of the choices of contours appearing in the definition of each term for $\epsilon_j^\pm\in(0,2\pi)$, real, and expressible as a Fredholm determinant.\\\\
The terms in the series take the form,
\begin{multline}
    \mathcal{G}^{(m)}(\kappa) =\frac{1}{m!}\prod_{j=1}^m\int_{\mathcal{C}_+}\frac{\dd \lambda_j}{2\pi}\,\frac{1}{m!}\prod_{j=1}^m\int_{\mathcal{C}_-}\frac{\dd \sigma_j}{2\pi}\,\frac{\prod_{1\leq i<j\leq m}(\lambda_i-\lambda_j)^2(\sigma_i-\sigma_j)^2}{\prod_{1\leq i,j\leq m}(\lambda_i-\sigma_j)^2}\\
    \times \exp\left[4i\kappa\sum_{j=1}^m\left(\sinh(\lambda_j)-\sinh(\sigma_j)\right)\right]\,.\label{eqn:mathcalGmkappa_SJT}
\end{multline}
Once again, we are taking the contours $\mathcal{C}_\pm$ to be above and below the real axis, such that
\begin{equation}
    \lambda_j=x_j+\frac{i\epsilon_j^+}{2}\,,\qquad\qquad\sigma_j=y_j-\frac{i\epsilon_j^-}{2}\,,
\end{equation}
with $x_j, \, y_j\in\mathbb{R}$ and $\epsilon_j^\pm\in(0,2\pi)$. Using the Cauchy determinant identity, the terms can be written as,
\begin{multline}
    \mathcal{G}^{(m)}(\kappa)=\sum_{p,q\,\in\,S_m}\text{sgn}(p)\,\text{sgn}(q)\frac{1}{m!}\prod_{j=1}^m\int_{-\infty}^{\infty}\frac{\dd x_j}{2\pi}\,\frac{1}{m!}\prod_{j=1}^m\int_{-\infty}^{\infty}\frac{\dd y_j}{2\pi}\,\prod_{j=1}^m\frac{1}{x_j-y_{p(j)}+\frac{i}{2}(\epsilon_j^++\epsilon_{p(j)}^-)} \\
    \times\prod_{j=1}^m\frac{1}{x_j-y_{q(j)}+\frac{i}{2}(\epsilon_j^++\epsilon_{q(j)}^-)}\times\prod_{j=1}^m\exp\left[4i\kappa\; \sinh\left(x_j+\frac{i\epsilon_j^+}{2}\right)-4i\kappa\; \sinh\left(y_j-\frac{i\epsilon_j^-}{2}\right)\right]\,.
\label{eqn:ToBeAnalyzedSJT}
\end{multline}
The real part of the exponent is
\begin{equation}
    \text{Re}\left[4i\kappa\sum_{j=1}^m\left(\sinh(\lambda_j)-\sinh(\sigma_j)\right)\right]=-4\kappa\sum_{j=1}^m\left(\sin\left(\frac{\epsilon_j^+}{2}\right)\cosh(x_j)+\sin\left(\frac{\epsilon_j^-}{2}\right)\cosh(y_j)\right)\,,
\end{equation}
which means that it is strictly negative for $0<\epsilon_j^\pm<2\pi$. In fact, since $\cosh(u)>u^2/2$ for all $u\in\mathbb{R}$,
It follows that
\begin{equation}
    \left|\exp\left[4i\kappa\sum_{j=1}^m\left(\sinh(\lambda_j)-\sinh(\sigma_j)\right)\right]\right|<\exp\left[-2\kappa\sum_{j=1}^m\left(\sin\left(\frac{\epsilon_j^+}{2}\right)x_j^2+\sin\left(\frac{\epsilon_j^-}{2}\right)y_j^2\right)\right]\,.
\end{equation}
The denominators appearing in the integrand are also absolutely bounded as
\begin{equation}
    \left|\frac{1}{x_j-y_{p(j)}+\frac{i}{2}(\epsilon_j^++\epsilon_{p(j)}^-)}\frac{1}{x_j-y_{q(j)}+\frac{i}{2}(\epsilon_j^++\epsilon_{q(j)}^-)}\right|\leq\frac{4}{(\epsilon_j^++\epsilon_{p(j)}^-)(\epsilon_j^++\epsilon_{q(j)}^-)}\,.\label{eqn:denominator_bound_SJT}
\end{equation}
Thus, the integral (\ref{eqn:ToBeAnalyzedSJT}) is absolutely convergent for any choices of regularization parameters satisfying $0<\epsilon_j^\pm<2\pi$ since it is bounded by a product of Gaussian integrals.\\\\
Let us now show that (\ref{eqn:ToBeAnalyzedSJT}) is also independent of the choices of regularization parameters. The integrand is holomorphic in each of the variables $x_j$ and $y_j$, and we can once again employ a contour integral argument, similar to the one in section \ref{sec:3.1}. Once again, we need to show that the absolute value of the $x_j$ integral over the vertical sides of the rectangle of length $L$ and height $\Delta\epsilon_j^+$ vanishes as $L\rightarrow\infty$. The horizontal sides of the rectangle are located at heights $\epsilon_j^+$ and $\epsilon_j^++\Delta\epsilon_j^+$ above the real axis, satisfying $0<\epsilon_j^+<\epsilon_j^++\Delta\epsilon_j^+<2\pi$. We can use the bound in (\ref{eqn:denominator_bound_SJT}) for the denominator. For the exponential factor, let $\hat{\epsilon}$ be the choice between $\epsilon_j^+$ and $\epsilon_j^++\Delta\epsilon_j^+$ which minimizes $\sin(\hat{\epsilon}/2)$. We can then bound the absolute value of the exponential factor appearing in the integrand on the vertical sides as,
\begin{equation}
    \left|\exp\left[4i\kappa\,\sinh(\lambda_j)\right]\right|=\exp\left[-4\kappa\,\sin\left(\frac{u+\epsilon_j^+}{2}\right)\cosh(L)\right]<\exp\left[-4\kappa\sin\left(\frac{\hat{\epsilon}}{2}\right)\cosh(L)\right]\,.
\end{equation}
Thus, as $L\rightarrow\infty$, the integrand vanishes on the vertical sides of the rectangle, which implies that the integrals along the horizontal sides located at height $\epsilon_j^+$ and at height $\epsilon_j^++\Delta\epsilon_j^+$ are equal to one another. This argument breaks down if we choose $\epsilon_j^++\Delta\epsilon_j^+>2\pi$, because $\sin(\hat{\epsilon}/2)$ can now be negative. A similar argument shows that the integral (\ref{eqn:ToBeAnalyzedSJT}) is independent of the regularization parameters $\epsilon_j^-$.\\\\
In the following, we will choose all regularization parameters to be equal to one another, $\epsilon_j^+=\epsilon_j^-=\epsilon$, such that
\begin{multline}
    \mathcal{G}^{(m)}(\kappa)=\sum_{p,q\,\in\,S_m}\text{sgn}(p)\,\text{sgn}(q)\frac{1}{m!}\prod_{j=1}^m\int_{-\infty}^{\infty}\frac{\dd x_j}{2\pi}\,\frac{1}{m!}\prod_{j=1}^m\int_{-\infty}^{\infty}\frac{\dd y_j}{2\pi}\,\prod_{j=1}^m\frac{1}{x_j-y_{p(j)}+i\epsilon}\\
    \times\prod_{j=1}^m\frac{1}{x_j-y_{q(j)}+i\epsilon}\prod_{j=1}^m\exp\left[4i\kappa\; \sinh\left(x_j+\frac{i\epsilon}{2}\right)-4i\kappa\; \sinh\left(y_j-\frac{i\epsilon}{2}\right)\right]\,.
\label{eqn:ToBeAnalyzedSJTequaleps}
\end{multline}
Taking the complex conjugate of (\ref{eqn:ToBeAnalyzedSJTequaleps}) and performing a relabeling of the integration variables, one recovers the same expression, which means that the terms $\mathcal{G}^{(m)}(\kappa)$ are real. Thus, the partition function $\mathcal{Z}(\kappa)$ given by (\ref{eqn:ZSJTexpression}) is real.\\\\
One can follow the argument in section \ref{sec:4} and define the integral operator $T$ on $L^2(\mathbb{R})$ with momentum space representation
\begin{equation}
    T\ket{p}=\int_{-\infty}^{+\infty}\dd p'\,f(p,p')\ket{p'}\,,
\end{equation}
where
\begin{equation}
    f(p,p')=\braket{p'|T|p}=\theta(p)\theta(p')\int_{\mathcal{C}_+}\frac{\dd\lambda}{2\pi}\int_{\mathcal{C}_-}\frac{\dd\sigma}{2\pi}\,e^{ip\lambda}e^{-ip'\sigma}\frac{i}{\lambda-\sigma}\exp\left[4i\kappa(\sinh(\lambda)-\sinh(\sigma))\right]\,.
\end{equation}
One can then show that $T$ is self-adjoint and trace-class since
\begin{equation}
    \|T\|_1=-\mathcal{G}^{(1)}(\kappa)<\infty\,,
\end{equation}
which allows us to construct its associated Fredholm determinant. Once again, we find
\begin{equation}
    \mathcal{G}^{(m)}(\kappa)=(-1)^m\,\Tr\,\Lambda^m(T)\,,
\end{equation}
from which we conclude that the partition function of JT supergravity can be written as the Fredholm determinant associated with the trace-class operator $T$,\footnote{Recall that the Fredholm determinant associated with a trace-class operator $T$ is well-defined (has a convergent expression) \cite{lax2014functional}, which means that the expression (\ref{eqn:ZSJTexpression}) we have found for the partition function $\mathcal{Z}(\kappa)$ is convergent.}
\begin{equation}
    \mathcal{Z}(\kappa)=\det\left(\mathbb{1}-T\right)\,.
\end{equation}

\bibliographystyle{apsrev4-1long}
\bibliography{biblio}

\end{document}